\documentclass[10pt]{article}
\usepackage{graphicx}
\usepackage{amsmath}
\usepackage{amssymb}
\usepackage{caption2}
\begin{document}
\bibliographystyle{prsty}
\begin{center}
{\large {\bf \sc{The singly-charmed pentaquark molecular states via the QCD sum rules }}} \\[2mm]
Qi Xin ${}^{*\dagger}$, Xiao-Song Yang ${}^{*\dagger}$,
Zhi-Gang  Wang ${}^*$ \footnote{E-mail: zgwang@aliyun.com.  }

Department of Physics, North China Electric Power University, Baoding 071003, P. R. China ${}^*$\\
School of Nuclear Science and Engineering, North China Electric Power University, Beijing 102206, P. R. China ${}^\dagger$
\end{center}

\begin{abstract}
In this work, we systematically investigate the singly-charmed pentaquark molecular states $D^{(*)}N$, $D^{(*)}\Xi^{(*)}$ and $D_s^{(*)}\Xi^{(*)}$ with the QCD sum rules by carrying out the operator product expansion up to the vacuum condensates of dimension 13 and taking fully  account of the light-flavor $SU(3)$ breaking effects. The numerical results favor assigning the $\Omega_c(3185)$ as the $D\Xi$ molecular state  with the $J^P=\frac{1}{2}^-$ and $| I,I_3 \rangle=| 0,0 \rangle$,  assigning the $\Omega_c(3327)$ as the $D^*\Xi$ molecular state with the $J^P=\frac{3}{2}^-$ and $|I,I_3 \rangle=| 0,0 \rangle$,   assigning the $\Sigma_c(2800)$   as the $DN$ molecular state with the $J^P=\frac{1}{2}^-$ and $| I,I_3 \rangle=| 1,0 \rangle$,  and assigning the $\Lambda_c(2940)/\Lambda_c(2910)$ as the $D^*N$ molecular state with the $J^P=\frac{3}{2}^-$ and $| I,I_3 \rangle=| 0,0 \rangle$.   Other potential  molecule candidates are also   predicted, which  may be observed in future experiments. For example, we can search for the $D\Xi$ and $D^*\Xi$ molecular states with the isospin $| I,I_3 \rangle=| 1,\pm1 \,\rangle$  in  the $\Xi_c^+\bar{K}^0$  and $\Xi_c^0\bar{K}^-$ mass spectrum respectively in the future, which could shed light on the nature of the $\Omega_c(3185/3327)$.
\end{abstract}



\section{Introduction}
Recently, two new excited states $\Omega_c(3185)^0$ and $\Omega_c(3327)^0$ were observed by the LHCb collaboration in the  $\Xi_c^+K^-$ invariant mass distribution \cite{LHCb2023-omegac}, they lie  near the thresholds of the meson-baryon pairs $D\Xi$ and $D^*\Xi$, respectively,  the measured Breit-Wigner masses and decay widths are
\begin{flalign}
& M_{\Omega_c(3185)} = 3185.1\pm1.7^{+7.4}_{-0.9}\pm0.2 \mbox{ MeV}\,,\,\Gamma_{\Omega_c(3185)} = 50\pm7 ^{+10}_{-20}\mbox{ MeV} \,,\nonumber \\
& M_{\Omega_c(3327)} = 3327.1\pm1.2^{+0.1}_{-1.3} \pm0.2\mbox{ MeV}\,,\,\Gamma_{\Omega_c(3327)} = 20\pm5 ^{+13}_{-1}\mbox{ MeV} \, .
\end{flalign}
 Prior to this, the LHCb collaboration announced five narrow structures named $\Omega_c(3000)^0$, $\Omega_c(3050)^0$, $\Omega_c(3066)^0$, $\Omega_c(3090)^0$ and $\Omega_c(3119)^0$ in 2017 \cite{LHCb2017-omegac}. Later, the Belle collaboration confirmed four of them, $\Omega_c(3000)^0$, $\Omega_c(3050)^0$, $\Omega_c(3066)^0$ and  $\Omega_c(3090)^0$, in the $\Xi_c^+K^-$ decay mode in the $e^+ e^-$ collisions \cite{Belle2017-omegac}.

As early as 2005, the Belle collaboration tentatively identified the $\Sigma_c(2800)^{0}$, $\Sigma_c(2800)^{+}$ and $\Sigma_c(2800)^{++}$ as isospin triplet states in the $\Lambda_c^+\pi^{-/0/+}$ mass spectrum, and provisionally designated the particles with the spin-parity $J^P=\frac{3}{2}^-$ based on the measured masses \cite{Belle-2005-sig2800}, the measured mass differences and decay widths are
\begin{flalign}
& M_{\Sigma_c(2800)^{0}} - M_{\Lambda^+}=515.4^{+3.2}_{-3.1}{}^{+2.1}_{-6.0}\, \mbox{MeV}\,,\,\Gamma_{\Sigma_c(2800)^{0}}=61^{+18}_{-13}{}^{+22}_{-13} \mbox{ MeV}\,,\nonumber \\
& M_{\Sigma_c(2800)^{+}} - M_{\Lambda^+}=505.4^{+5.8}_{-4.6}{}^{+12.4}_{-2.0} \,\mbox{MeV}\,,\,\Gamma_{\Sigma_c(2800)^{+}}=62^{+37}_{-23}{}^{+52}_{-38} \mbox{ MeV}\,,\nonumber \\
& M_{\Sigma_c(2800)^{++}} - M_{\Lambda^+}= 514.5^{+3.4}_{-3.1}{}^{+2.8}_{-4.9} \,\mbox{MeV}\,,\,\Gamma_{\Sigma_c(2800)^{++}}=75^{+18}_{-13}{}^{+12}_{-11}\mbox{ MeV}\, .
\end{flalign}
 In 2008, the BaBar collaboration observed the $\Sigma_c(2800)^{0}$ in the $\Lambda_c^+\pi^-$ mass spectrum  with the possible spin-parity  $J^P=\frac{1}{2}^- $ \cite{BaBar-2008-sig2800}, the mass and decay width are determined to be,
\begin{flalign}
& M_{\Sigma_c(2800)} = 2846\pm8\pm10 \mbox{ MeV}\,,\,\Gamma_{\Sigma_c(2800)} = 86{}^{+33}_{-22}\pm7\mbox{ MeV} \, .
\end{flalign}
So far, the experimental data cannot notarize the spin-parity $J^P$ of the $\Sigma_c(2800)^0$.

In 2007, the BaBar collaboration searched for charmed baryons in the $D^0p$ invariant mass spectrum and found the $\Lambda_c(2940)$ \cite{BaBar-2940-2007}. Subsequently, the Belle collaboration  verified the $\Lambda_c(2940)$ in the decay mode $\Lambda_c(2940) \!\!\to\!\Sigma_c(2455) \pi$ \cite{Belle-2940-2007}. In 2017, the spin-parity $J^P=\frac{3}{2}^-$ of the $\Lambda(2940)^+$ resonances were measured by the amplitude analysis of the process $\Lambda_b^0 \!\!\to\!D^0p \pi^-$ \cite{LHCb-2017-2940}. The measured masses and decay widths for the three experiments are listed as follows,
\begin{flalign}
& M_{\Lambda_c(2940)} = 2939.8 \pm1.3 \pm1.0 \mbox{ MeV}\,,\,\Gamma_{\Lambda_c(2940)} = 17.5\pm5.0\pm5.9 \mbox{ MeV} \,\,\,\,\,\,{\rm(BaBar)}\, , \nonumber \\
& M_{\Lambda_c(2940)} = 2938.0 \pm1.3 ^{+2.0}_{-4.0} \mbox{ MeV}\,,\,\Gamma_{\Lambda_c(2940)} = 13{}^{+8}_{-5}{}^{+27}_{-7} \mbox{ MeV} \,\,\,\,\,\,{\rm(Belle)}\, , \nonumber \\
& M_{\Lambda_c(2940)} = 2944.8 ^{+3.5}_{-2.5} \pm0.4 {}^{+0.1}_{-4.6} \mbox{ MeV}\,,\,\Gamma_{\Lambda_c(2940)} = 27.7^{+8.2}_{-6.0}\pm0.9{}^{+5.2}_{-10.4} \mbox{ MeV} \,\,\,\,\,\,{\rm(LHCb)}\,.
\end{flalign}
Last year, the Belle collaboration explored the decays $\bar{B}^0 \!\!\to\!\Sigma_c(2455)^{0,++} \pi^{\pm}\bar{p}$ and found a new structure in the $\Sigma_c(2455)^{0,++}\pi^{\pm}$ mass spectrum with a significance of $4.2\sigma$ including systematic uncertainty and named it as $\Lambda_c(2910)^+$ \cite{Belle2022-2910}, the mass and decay width are measured to be,
\begin{flalign}
& M_{\Lambda_c(2910)} = 2913.8 \pm5.6 \pm3.8 \mbox{ Mev}\,,\,\Gamma_{\Lambda_c(2910)} = 51.8\pm20.0\pm18.8 \mbox{ MeV} \,\,.
\end{flalign}
The $\Sigma_c(2800)$, $\Lambda_c(2940)$ and $\Lambda_c(2910)$ lie near the  $D^{(*)}N$ thresholds, which gives us the inspiration that they might be the $D^{(*)}N$ molecular states and it is interesting to study the singly-charmed pentaquark molecules.

The experimental discoveries served as bases for several theoretical studies, the theorists  attempted to explain the aforementioned excited $\Sigma_c$, $\Lambda_c$, and $\Omega_c$ baryon states and explore the internal structures. One of the goals is attempting  to interpret those baryons as charmed pentaquark molecular states, for example, assigning  the $\Sigma_c(2800)$ as the $DN$ molecular (bound) state \cite{2800-GFK,PJL-2800-2940-1,GXH-2800,ZD-2800-2940,Valderrama-2800-2940-3185-3327}, and  assigning the $\Lambda_c(2940)$ as the $D^*N$  molecular state \cite{PJL-2800-2940-1,PJL-2800-2940-2,ZD-2800-2940,Valderrama-2800-2940-3185-3327,
ZLZ-2910-2940,ZSL-2940,HY-2940}, although there remains controversy over its quantum numbers $J^P$.  However, there are some dissenting opinions, it is seemly impossible to explain the $\Sigma_c(2800)$ ($\Lambda_c(2940)$)  in terms of the $DN$ ($D^*N$) molecular states in some theoretical approaches, for example,  the potential quark model \cite{PJL-2800-2940-2}, the QCD sum rules \cite{ZJR-siglam} (the chiral effective field theory \cite{ZSL-2940}). While the  $D\Xi$ and $D^* \Xi$  systems could form  bound states or scattering states \cite{GXH-DXi,LMZ-DXi}, and  the $\Omega_c(3185)^0$  and $\Omega_c(3327)^0$ can be assigned as the $D\Xi$ and $D^* \Xi$ molecular states, respectively  \cite{Valderrama-2800-2940-3185-3327,FJW-3185-3327}.

The QCD sum rules are expected to give a bright solution for the masses and decay widths of the pentaquark states, just like what they have done  in  previous works, such as the singly-charmed pentaquark states \cite{ZGWJXZ-omegac-penta-mole,WDW-omegac-penta-mole,Charm-Albuquerque}, hidden-charm  pentaquark (molecular) states
\cite{ZGW-Pc4380-penta,ZGWTH-penta-1/2,ZWD-penta-3/2,ZGW-Pc4312-penta,
ZGW-Pcs4459-penta,Penta-Azizi-2021,Penta-Azizi-2022,Penta-ChenHX}
(\cite{ZGW-DSigmac-penta-mole,ZGWQX-DXic-penta-mole,WWYX-DSigmac-penta-mole-iso,
XWWZGW-DXic-penta-mole-iso,XWWZGW-Pcs4338-penta-mole,Penta-mole-ZhangJR,Penta-mole-CHX-2019,
Penta-mole-CHX-2021,Penta-mole-Azizi-2017,Penta-mole-LiuYL-2020,Penta-mole-WZG-decay}), and doubly-charmed pentaquark states \cite{ZGW-doub-charmed-penta}. For the electromagnetic
 properties of the hidden-charm pentaquark (molecular) states from the QCD sum rules, one can consult Refs.\cite{Penta-mole-EM-HuangMQ,Penta-EM-Qzdem}.
Based on our previous research foundation on the pentaquark masses \cite{ZGWJXZ-omegac-penta-mole,WDW-omegac-penta-mole,ZGW-Pc4380-penta,ZGWTH-penta-1/2,ZWD-penta-3/2,
ZGW-Pc4312-penta,ZGW-Pcs4459-penta,ZGW-DSigmac-penta-mole,ZGWQX-DXic-penta-mole,
WWYX-DSigmac-penta-mole-iso,XWWZGW-DXic-penta-mole-iso,XWWZGW-Pcs4338-penta-mole,ZGW-doub-charmed-penta}, we take those excited charmed baryon states as the pentaquark molecular states, and  probe the nature of the singly-charmed baryon family, it is the major pursuit of the present work.

The paper is arranged  as follows: we study the charmed pentaquark molecular states with different isospins using the QCD sum rules in section 2; the numerical results and discussions are presented in section 3; section 4 is reserved for conclusions.

\section{QCD sum rules for the singly-charmed pentaquark molecular states}

The two-point correlation functions $\Pi(p)$, $\Pi_{\mu\nu}(p)$
and $\Pi_{\mu\nu\alpha\beta}(p)$ in the QCD sum rules can be written in the following form,

\begin{eqnarray}
\Pi(p)&=&i\int d^4x e^{ip \cdot x} \langle0|T\left\{J(x)\bar{J}(0)\right
\}|0\rangle \, , \nonumber\\
\Pi_{\mu\nu}(p)&=&i\int d^4x e^{ip \cdot x} \langle0|T\left\{J_\mu(x)\bar{J}_\nu(0)\right
\}|0\rangle \, , \nonumber \\
\Pi_{\mu\nu\alpha\beta}(p)&=&i\int d^4x e^{ip \cdot x} \langle0|T\left\{J_{\mu\nu}(x)
\bar{J}_{\alpha\beta}(0)\right\}|0\rangle \, ,
\end{eqnarray}
where the interpolating currents,
\begin{eqnarray}
J(x)&=&J^{DN}_{\mid 1,0 \rangle}(x),\, J^{DN}_{\mid 0,0 \rangle}(x),\, J^{D\Xi}_{\mid 1,0 \rangle}(x),\, J^{D\Xi}_{\mid 0,0 \rangle}(x), \,J^{D_s\Xi}_{\mid \frac{1}{2},\pm\frac{1}{2} \rangle}(x) \,,\nonumber \\
J_\mu(x)&=&J^{D^{*}N}_{\mid 1,0 \rangle}(x),\, J^{D^{*}N}_{\mid 0,0 \rangle}(x), \, J^{D^{*}\Xi}_{\mid 1,0 \rangle}(x),\, J^{D^{*}\Xi}_{\mid 0,0 \rangle}(x), \,J^{D_s^*\Xi}_{\mid \frac{1}{2},\pm\frac{1}{2} \rangle}(x),\, J^{D\Xi^{*}}_{\mid 1,0 \rangle}(x),\, J^{D\Xi^{*}}_{\mid 0,0 \rangle}(x),\, J^{D_s\Xi^*}_{\mid \frac{1}{2},\pm\frac{1}{2} \rangle}(x)\,,\nonumber \\
J_{\mu\nu}(x)&=&J^{D^{*}\Xi^{*}}_{\mid 1,0 \rangle}(x), \,J^{D^{*}\Xi^{*}}_{\mid 0,0 \rangle}(x),\,  J^{D_s^*\Xi^*}_{\mid \frac{1}{2},\pm\frac{1}{2} \rangle}(x) \,  ,
\end{eqnarray}
\begin{eqnarray}\label{current-DN}
J^{DN}_{\mid 1,0 \rangle}(x) &=& \frac{1}{\sqrt{2}}  J^{D^0}_{\mid \frac{1}{2},-\frac{1}{2} \rangle}(x) J^{N^+}_{\mid \frac{1}{2},\frac{1}{2} \rangle}(x) +\frac{1}{\sqrt{2}}J^{D^+}_{\mid \frac{1}{2},\frac{1}{2} \rangle}(x)J^{N^0}_{\mid \frac{1}{2},-\frac{1}{2} \rangle}  (x)  \,,\nonumber \\
J^{DN}_{\mid 0,0 \rangle}(x) &=& \frac{1}{\sqrt{2}}  J^{D^0}_{\mid \frac{1}{2},-\frac{1}{2} \rangle}(x) J^{N^+}_{\mid \frac{1}{2},\frac{1}{2} \rangle}(x)        -\frac{1}{\sqrt{2}}J^{D^+}_{\mid \frac{1}{2},\frac{1}{2} \rangle}(x)J^{N^0}_{\mid \frac{1}{2},-\frac{1}{2} \rangle}  (x)       \,,\nonumber \\
J^{D\Xi}_{\mid 1,0 \rangle}(x) &=& \frac{1}{\sqrt{2}}J^{D^0}_{\mid \frac{1}{2},-\frac{1}{2} \rangle}(x) J^{\Xi^0}_{\mid \frac{1}{2},\frac{1}{2} \rangle}(x)+\frac{1}{\sqrt{2}} J^{D^+}_{\mid \frac{1}{2},\frac{1}{2} \rangle}(x)J^{\Xi^-}_{\mid \frac{1}{2},-\frac{1}{2} \rangle}(x)    \,,\nonumber \\
J^{D\Xi}_{\mid 0,0 \rangle}(x) &=& \frac{1}{\sqrt{2}}J^{D^0}_{\mid \frac{1}{2},-\frac{1}{2} \rangle}(x) J^{\Xi^0}_{\mid \frac{1}{2},\frac{1}{2} \rangle}(x)-\frac{1}{\sqrt{2}} J^{D^+}_{\mid \frac{1}{2},\frac{1}{2} \rangle}(x)J^{\Xi^-}_{\mid \frac{1}{2},-\frac{1}{2} \rangle} (x)          \,,\nonumber \\
J^{D_s\Xi}_{\mid \frac{1}{2},\pm\frac{1}{2} \rangle}(x) &=& J^{D_s^+}_{\mid 0,0 \rangle} (x) J^{\Xi^{0/-}} _{\mid \frac{1}{2},\pm\frac{1}{2} \rangle} (x)       \, ,
\end{eqnarray}

\begin{eqnarray}\label{current-DN-A}
J^{D^{*}N}_{\mid 1,0 \rangle}(x) &=& \frac{1}{\sqrt{2}}  J^{D^{*0}}_{\mid \frac{1}{2},-\frac{1}{2} \rangle}(x) J^{N^+}_{\mid \frac{1}{2},\frac{1}{2} \rangle}(x)         +\frac{1}{\sqrt{2}}J^{D^{*+}}_{\mid \frac{1}{2},\frac{1}{2} \rangle}(x)J^{N^0}_{\mid \frac{1}{2},-\frac{1}{2} \rangle} (x)\,,\nonumber \\
J^{D^{*}N}_{\mid 0,0 \rangle}(x) &=& \frac{1}{\sqrt{2}}  J^{D^{*0}}_{\mid \frac{1}{2},-\frac{1}{2} \rangle}(x) J^{N^+}_{\mid \frac{1}{2},\frac{1}{2} \rangle}(x)         -\frac{1}{\sqrt{2}}J^{D^{*+}}_{\mid \frac{1}{2},\frac{1}{2} \rangle}(x)J^{N^0}_{\mid \frac{1}{2},-\frac{1}{2} \rangle} (x)\,,\nonumber \\
J^{D^{*}\Xi}_{\mid 1,0 \rangle}(x) &=& \frac{1}{\sqrt{2}}J^{D^{*0}}_{\mid \frac{1}{2},-\frac{1}{2} \rangle}(x) J^{\Xi^0}_{\mid \frac{1}{2},\frac{1}{2} \rangle}(x)+\frac{1}{\sqrt{2}} J^{D^{*+}}_{\mid \frac{1}{2},\frac{1}{2} \rangle}(x)J^{\Xi^-}_{\mid \frac{1}{2},-\frac{1}{2} \rangle} (x)          \,,\nonumber \\
J^{D^{*}\Xi}_{\mid 0,0 \rangle}(x) &=& \frac{1}{\sqrt{2}}J^{D^{*0}}_{\mid \frac{1}{2},-\frac{1}{2} \rangle}(x) J^{\Xi^0}_{\mid \frac{1}{2},\frac{1}{2} \rangle}(x)-\frac{1}{\sqrt{2}} J^{D^{*+}}_{\mid \frac{1}{2},\frac{1}{2} \rangle}(x)J^{\Xi^-}_{\mid \frac{1}{2},-\frac{1}{2} \rangle} (x)              \,, \nonumber \\
J^{D_s^*\Xi}_{\mid \frac{1}{2},\pm\frac{1}{2} \rangle}(x) &=& J^{D_s^{*+}}_{\mid 0,0\rangle} (x)  J^{\Xi^{0/-}}_{\mid \frac{1}{2},\pm\frac{1}{2} \rangle} (x)       \, ,      \nonumber \\
J^{D\Xi^{*}}_{\mid 1,0 \rangle}(x) &=& \frac{1}{\sqrt{2}}J^{D^{0}}_{\mid \frac{1}{2},-\frac{1}{2} \rangle}(x) J^{\Xi^{*0}}_{\mid \frac{1}{2},\frac{1}{2} \rangle}(x)+\frac{1}{\sqrt{2}} J^{D^{+}}_{\mid \frac{1}{2},\frac{1}{2} \rangle}(x)J^{\Xi^{*-}}_{\mid \frac{1}{2},-\frac{1}{2} \rangle} (x)          \,,\nonumber \\
J^{D\Xi^{*}}_{\mid 0,0 \rangle}(x) &=& \frac{1}{\sqrt{2}}J^{D^{0}}_{\mid \frac{1}{2},-\frac{1}{2} \rangle}(x) J^{\Xi^{*0}}_{\mid \frac{1}{2},\frac{1}{2} \rangle}(x)-\frac{1}{\sqrt{2}} J^{D^{+}}_{\mid \frac{1}{2},\frac{1}{2} \rangle}(x)J^{\Xi^{*-}}_{\mid \frac{1}{2},-\frac{1}{2} \rangle} (x)            \,, \nonumber \\
J^{D_s\Xi^*}_{\mid \frac{1}{2},\pm\frac{1}{2} \rangle}(x) &=& J^{D_s^{+}}_{\mid 0,0\rangle} (x)  J^{\Xi^{*0/-}}_{\mid \frac{1}{2},\pm\frac{1}{2} \rangle} (x)       \, ,
\end{eqnarray}

\begin{eqnarray}\label{current-DN-T}
J^{D^{*}\Xi^{*}}_{\mid 1,0 \rangle}(x) &=& \frac{1}{\sqrt{2}}J^{D^{*0}}_{\mid \frac{1}{2},-\frac{1}{2} \rangle}(x) J^{\Xi^{*0}}_{\mid \frac{1}{2},\frac{1}{2} \rangle}(x)+\frac{1}{\sqrt{2}} J^{D^{*+}}_{\mid \frac{1}{2},\frac{1}{2} \rangle}(x)J^{\Xi^{*-}}_{\mid \frac{1}{2},-\frac{1}{2} \rangle} (x)          \,,\nonumber \\
J^{D^{*}\Xi^{*}}_{\mid 0,0 \rangle}(x) &=& \frac{1}{\sqrt{2}}J^{D^{*0}}_{\mid \frac{1}{2},-\frac{1}{2} \rangle}(x) J^{\Xi^{*0}}_{\mid \frac{1}{2},\frac{1}{2} \rangle}(x)-\frac{1}{\sqrt{2}} J^{D^{*+}}_{\mid \frac{1}{2},\frac{1}{2} \rangle}(x)J^{\Xi^{*-}}_{\mid \frac{1}{2},-\frac{1}{2} \rangle} (x)            \,, \nonumber \\
J^{D_s^*\Xi^*}_{\mid \frac{1}{2},\pm\frac{1}{2} \rangle}(x) &=& J^{D_s^{*+}}_{\mid 0,0\rangle} (x)  J^{\Xi^{*0/-}}_{\mid \frac{1}{2},\pm\frac{1}{2} \rangle} (x)       \, ,
\end{eqnarray}
and
\begin{eqnarray}
J^{D^0}_{\mid \frac{1}{2},-\frac{1}{2} \rangle}(x) &=& \bar{u}(x)i\gamma_5 c(x)\, ,      \nonumber \\
J^{D^+}_{\mid \frac{1}{2},\frac{1}{2} \rangle}(x) &=& -\bar{d}(x)i\gamma_5 c(x)\, ,      \nonumber \\
J^{D_s^+}_{\mid 0,0 \rangle}(x) &=& \bar{s}(x)i\gamma_5 c(x)\, ,      \nonumber \\
J^{D^{*0}}_{\mid \frac{1}{2},-\frac{1}{2} \rangle}(x) &=& \bar{u}(x)\gamma_\mu c(x)\, ,      \nonumber \\
J^{D^{*+}}_{\mid \frac{1}{2},\frac{1}{2} \rangle}(x) &=& -\bar{d}(x)\gamma_\mu c(x)\, ,      \nonumber \\
J^{D_s^{*+}}_{\mid 0,0 \rangle}(x) &=& \bar{s}(x)\gamma_\mu c(x)\, ,
\end{eqnarray}

\begin{eqnarray}
J^{N^+}_{\mid \frac{1}{2},\frac{1}{2} \rangle}(x) &=&\varepsilon^{ijk}u_i^T(x)C\gamma_\mu u_j(x)\gamma^\mu\gamma_5 d_k(x)\, ,      \nonumber \\
J^{N^0}_{\mid \frac{1}{2},-\frac{1}{2} \rangle}(x) &=&\varepsilon^{ijk}d_i^T(x)C\gamma_\mu d_j(x)\gamma^\mu\gamma_5 u_k(x)\, ,      \nonumber \\
J^{\Xi^0}_{\mid \frac{1}{2},\frac{1}{2} \rangle}(x) &=&\varepsilon^{ijk}s_i^T(x)C\gamma_\mu s_j(x)\gamma^\mu\gamma_5 u_k(x)\, ,      \nonumber \\
J^{\Xi^-}_{\mid \frac{1}{2},-\frac{1}{2} \rangle}(x) &=&\varepsilon^{ijk}s_i^T(x)C\gamma_\mu s_j(x)\gamma^\mu\gamma_5 d_k(x)\, ,      \nonumber \\
J^{\Xi^{*0}}_{\mid \frac{1}{2},\frac{1}{2} \rangle}(x) &=&\varepsilon^{ijk}s_i^T(x)C\gamma_\mu s_j(x) u_k(x)\, ,      \nonumber \\
J^{\Xi^{*-}}_{\mid \frac{1}{2},-\frac{1}{2} \rangle}(x) &=&\varepsilon^{ijk}s_i^T(x)C\gamma_\mu s_j(x) d_k(x)\, ,
\end{eqnarray}
the $i$, $j$, $k$ are color indexes, the subscripts  $| 1,0 \rangle$, $|0,0 \rangle$,
$|\frac{1}{2},\frac{1}{2} \rangle$ and $|\frac{1}{2},-\frac{1}{2} \rangle$ are isospin indexes ${| I,I_3 \rangle}$. The $J^{D^0}(x)$, $J^{D^+}(x)$, $J^{D_s^+}(x)$, $J^{D^{*0}}(x)$, $J^{D^{*+}}(x)$ and $J^{D_s^{*+}}(x)$ are the standard currents interpolating  the mesons, and the $J^{N^+}(x)$, $J^{N^0}(x)$, $J^{\Xi^0}(x)$, $J^{\Xi^-}(x)$, $J^{\Xi^{*0}}(x)$ and $J^{\Xi^{*-}}(x)$ are the standard currents interpolating the baryons, where we have chosen the convention $|\frac{1}{2},\frac{1}{2} \rangle=-\bar{d}$ for the isospin eigenstate.
The currents $J(x)$  interpolate the $DN$, $D\Xi$ and $D_s\Xi$ pentaquark molecular states with the spin-parity $J^P={\frac{1}{2}}^-$, the currents $J_{\mu}(x)$  interpolate the $D^{*}N$, $D^{*}\Xi$, $D_s^*\Xi$, $D\Xi^{*}$ and $D_s\Xi^{*}$ pentaquark molecular states with the spin-parity $J^P={\frac{3}{2}}^-$, and the currents $J_{\mu\nu}(x)$  interpolate the $D^{*}\Xi^{*}$ and $D_s^*\Xi^{*}$ pentaquark molecular states with the spin-parity $J^P={\frac{5}{2}}^-$, where the  $DN$, $D\Xi$ and $D_s\Xi$ $\cdots$ represent the two color-neutral clusters having  the same quantum numbers as the $D$, $N$, $\Xi$, $D_s$ $\cdots$, they are not necessary to have the same masses as the physical mesons and baryons, as we take the local five-quark currents.

The currents $J(0)$, $J_\mu(0)$ and $J_{\mu\nu}(0)$ couple potentially to the $J^P={\frac{1}{2}}^\mp$, ${\frac{3}{2}}^\mp$ and ${\frac{5}{2}}^\mp$  singly-charmed  pentaquark molecular  states $P_{\frac{1}{2}}^\mp$, $P_{\frac{3}{2}}^\mp$, and $P_{\frac{5}{2}}^\mp$, respectively,
\begin{eqnarray}\label{J-lamda}
\langle 0| J (0)|P_{\frac{1}{2}}^{-}(p)\rangle &=&\lambda^{-}_{\frac{1}{2}} U^{-}(p,s) \, ,\nonumber  \\
\langle 0| J (0)|P_{\frac{1}{2}}^{+}(p)\rangle &=&\lambda^{+}_{\frac{1}{2}}i\gamma_5 U^{+}(p,s) \, , \nonumber \\
\langle 0| J_{\mu} (0)|P_{\frac{3}{2}}^{-}(p)\rangle &=&\lambda^{-}_{\frac{3}{2}} U^{-}_\mu(p,s) \, , \nonumber \\
\langle 0| J_{\mu} (0)|P_{\frac{3}{2}}^{+}(p)\rangle &=&\lambda^{+}_{\frac{3}{2}}i\gamma_5 U^{+}_{\mu}(p,s) \, , \nonumber \\
\langle 0| J_{\mu\nu} (0)|P_{\frac{5}{2}}^{-}(p)\rangle &=&\lambda^{-}_{\frac{5}{2}} U^{-}_{\mu\nu}(p,s) \, , \nonumber\\
\langle 0| J_{\mu\nu} (0)|P_{\frac{5}{2}}^{+}(p)\rangle &=&\lambda^{+}_{\frac{5}{2}}i\gamma_5 U^{+}_{\mu\nu}(p,s) \, ,
\end{eqnarray}
where the $U^\pm(p,s)$ are the Dirac spinors, the $U^{\pm}_\mu(p,s)$ and $U^{\pm}_{\mu\nu}(p,s)$ are the Rarita-Schwinger spinors \cite{ZGW-Pc4380-penta,ZGWTH-penta-1/2,ZGW-Pc4312-penta,ZGW-DSigmac-penta-mole,
WangHuang1508-2,WangHuang1508-3,Chung82-1,Chung82-2,Oka96}, and the
$\lambda^{\mp}_{\frac{1}{2}}$, $\lambda^{\mp}_{\frac{3}{2}}$and $\lambda^{\mp}_{\frac{5}{2}}$ are the corresponding pole residues.

At the hadron side of the correlation functions $\Pi(p)$, $\Pi_{\mu\nu}(p)$ and $\Pi_{\mu\nu\alpha\beta}(p)$, we isolate the  ground state contributions from the singly-charmed  pentaquark molecular states with the spin-parity $J^P={\frac{1}{2}}^\mp$, ${\frac{3}{2}}^\mp$ and ${\frac{5}{2}}^\mp$ respectively without contaminations according to the current-hadron couplings shown in Eqs.\eqref{J-lamda}, and get the hadronic representation \cite{QCDSR-SVZ79,QCDSR-Reinders85},
\begin{eqnarray}
\Pi(p) & = & {\lambda^{-}_{\frac{1}{2}}}^2  {\!\not\!{p}+ M_{-} \over M_{-}^{2}-p^{2}  } +  {\lambda^{+}_{\frac{1}{2}}}^2  {\!\not\!{p}- M_{+} \over M_{+}^{2}-p^{2}  } +\cdots  \, ,\nonumber\\
  &=&\Pi_{\frac{1}{2}}^1(p^2)\!\not\!{p}+\Pi_{\frac{1}{2}}^0(p^2)\, ,
\end{eqnarray}

\begin{eqnarray}
\Pi_{\mu\nu}(p) & = & {\lambda^{-}_{\frac{3}{2}}}^2  {\!\not\!{p}+ M_{-} \over M_{-}^{2}-p^{2}  } \left(- g_{\mu\nu}+\frac{\gamma_\mu\gamma_\nu}{3}+\frac{2p_\mu p_\nu}{3p^2}-\frac{p_\mu\gamma_\nu-p_\nu \gamma_\mu}{3\sqrt{p^2}}\right)\nonumber\\
&&+ {\lambda^{+}_{\frac{3}{2}}}^2  {\!\not\!{p}- M_{+} \over M_{+}^{2}-p^{2}  } \left(- g_{\mu\nu}+\frac{\gamma_\mu\gamma_\nu}{3}+\frac{2p_\mu p_\nu}{3p^2}-\frac{p_\mu\gamma_\nu-p_\nu \gamma_\mu}{3\sqrt{p^2}}\right)   +\cdots  \, ,\nonumber\\
   &=&-\Pi_{\frac{3}{2}}^1(p^2)\!\not\!{p}\,g_{\mu\nu}-\Pi_{\frac{3}{2}}^0(p^2)\,g_{\mu\nu}+\cdots\, ,
\end{eqnarray}

\begin{eqnarray}
\Pi_{\mu\nu\alpha\beta}(p) & = & {\lambda^{+}_{\frac{5}{2}}}^2  {\!\not\!{p}+ M_{+} \over M_{+}^{2}-p^{2}  } \left[\frac{ \widetilde{g}_{\mu\alpha}\widetilde{g}_{\nu\beta}+\widetilde{g}_{\mu\beta}\widetilde{g}_{\nu\alpha}}{2}-\frac{\widetilde{g}_{\mu\nu}\widetilde{g}_{\alpha\beta}}{5}
-\frac{1}{10}\left( \gamma_{\alpha}\gamma_{\mu}+\cdots\right)\widetilde{g}_{\nu\beta}
+\cdots\right]\nonumber\\
&&+   {\lambda^{-}_{\frac{5}{2}}}^2  {\!\not\!{p}- M_{-} \over M_{-}^{2}-p^{2}  }  \left[\frac{ \widetilde{g}_{\mu\alpha}\widetilde{g}_{\nu\beta}+\widetilde{g}_{\mu\beta}\widetilde{g}_{\nu\alpha}}{2}-\frac{\widetilde{g}_{\mu\nu}\widetilde{g}_{\alpha\beta}}{5}
+\cdots\right]   +\cdots \, , \nonumber\\
&=&\Pi_{\frac{5}{2}}^1(p^2)\!\not\!{p}\frac{g_{\mu\alpha}g_{\nu\beta}+g_{\mu\beta}g_{\nu\alpha}}{2} +\Pi_{\frac{5}{2}}^0(p^2)\,\frac{g_{\mu\alpha}g_{\nu\beta}+g_{\mu\beta}g_{\nu\alpha}}{2} + \cdots \, ,
\end{eqnarray}
where $\widetilde{g}_{\mu\nu}=g_{\mu\nu}-\frac{p_{\mu}p_{\nu}}{p^2}$
and $p^2=M^2_{\pm}$ on  mass-shell. We choose the components $\Pi_{\frac{1}{2}}^1(p^2)$, $\Pi_{\frac{1}{2}}^0(p^2)$, $\Pi_{\frac{3}{2}}^1(p^2)$, $\Pi_{\frac{3}{2}}^0(p^2)$, $\Pi_{\frac{5}{2}}^1(p^2)$ and $\Pi_{\frac{5}{2}}^0(p^2)$ to explore the spin-parity $J^{P}={\frac{1}{2}}^-$, ${\frac{3}{2}}^-$ and ${\frac{5}{2}}^-$ states, respectively.

Then we obtain the spectral densities at hadron side through dispersion relation,
\begin{eqnarray}
\frac{{\rm Im}\Pi_{j}^1(s)}{\pi}&=& {\lambda^{-}_{j}}^2 \delta\left(s-M_{-}^2\right)+{\lambda^{+}_{j}}^2 \delta\left(s-M_{+}^2\right) =\, \rho^1_{j,H}(s) \, ,\nonumber \\
\frac{{\rm Im}\Pi^0_{j}(s)}{\pi}&=&M_{-}{\lambda^{-}_{j}}^2 \delta\left(s-M_{-}^2\right)-M_{+}{\lambda^{+}_{j}}^2 \delta\left(s-M_{+}^2\right)
=\rho^0_{j,H}(s) \, ,
\end{eqnarray}
where $j=\frac{1}{2}$, $\frac{3}{2}$, $\frac{5}{2}$, the subscript $H$ denotes  the hadron side,
then we introduce the  weight functions $\sqrt{s}\exp\left(-\frac{s}{T^2}\right)$ and $\exp\left(-\frac{s}{T^2}\right)$ to obtain the QCD sum rules at the hadron side,
\begin{eqnarray}\label{QCDSR-M}
\int_{m_c^2}^{s_0}ds \left[\sqrt{s}\rho^1_{j,H}(s)+\rho^0_{j,H}(s)\right]\exp\left( -\frac{s}{T^2}\right)
&=&2M_{-}{\lambda^{-}_{j}}^2\exp\left( -\frac{M_{-}^2}{T^2}\right) \, ,
\end{eqnarray}
where the $s_0$ are the continuum threshold parameters and the $T^2$ are the Borel parameters. On the other hand,
we can obtain  the QCD sum rules at the hadron side,
\begin{eqnarray}\label{QCDSR-M-P}
\int_{m_c^2}^{s_0}ds \left[\sqrt{s}\rho^1_{j,H}(s)-\rho^0_{j,H}(s)\right]\exp\left( -\frac{s}{T^2}\right)
&=&2M_{+}{\lambda^{+}_{j}}^2\exp\left( -\frac{M_{+}^2}{T^2}\right) \, ,
\end{eqnarray}
for the positive-parity molecular states.
We separate the  contributions  of the negative parity  molecular states from that of the positive parity molecular states unambiguously. In the QCD sum rules, see Eqs.\eqref{QCDSR-M}-\eqref{QCDSR-M-P}, there only exist contributions from the negative-parity or positive-parity molecular states.

At the QCD side, we contract  the correlation functions $\Pi(p)$, $\Pi_{\mu\nu}(p)$ and $\Pi_{\mu\nu\alpha\beta}(p)$ with the Wick's theorem, for example, the correlation functions $\Pi^{DN}(p)$ (for the $J^{DN}_{\mid 1,0 \rangle}(x)$), $\Pi^{D\Xi}(p)$ (for the $J^{D\Xi}_{\mid 0,0 \rangle}(x)$), $\Pi^{D^*N}_{\mu\nu}(p)$ (for the $J^{D^*N}_{\mid 0,0 \rangle}(x)$) and $\Pi^{D^*\Xi}_{\mu\nu}(p)$ (for the $J^{D^*\Xi}_{\mid 0,0 \rangle}(x)$) can be written as
\begin{eqnarray}
\Pi^{DN}(p)&=&i\varepsilon^{ijk}\varepsilon^{i^{\prime}j^{\prime}k^{\prime}}\int d^4x e^{ip \cdot x}\nonumber\\
&&\left\{\begin{array}{l}+{\rm Tr}\! \left\{i\gamma_5 C_{mm^{\prime}}(x)i\gamma_5 U_{m^{\prime}m}(-x)\right\} {\rm Tr}\!\left\{\gamma_\mu  U_{jj^{\prime}}(x)\gamma_\nu C U_{ii^{\prime}}^T(x)C \right\} \gamma_5 \gamma^\mu D_{kk^{\prime}}(x) \gamma^\nu\gamma_5\\
+{\rm Tr}\! \left\{i\gamma_5 C_{mm^{\prime}}(x)i\gamma_5 D_{m^{\prime}m}(-x)\right\} {\rm Tr}\!\left\{\gamma_\mu  D_{jj^{\prime}}(x)\gamma_\nu C D_{ii^{\prime}}^T(x)C \right\} \gamma_5 \gamma^\mu U_{kk^{\prime}}(x) \gamma^\nu\gamma_5
\end{array}\right\} \, ,\nonumber\\
\end{eqnarray}
\begin{eqnarray}
\Pi^{D\Xi}(p)&=&i\varepsilon^{ijk}\varepsilon^{i^{\prime}j^{\prime}k^{\prime}}\int d^4x e^{ip \cdot x}\nonumber\\
&&\left\{\begin{array}{l}+{\rm Tr}\! \left\{i\gamma_5 C_{mm^{\prime}}(x)i\gamma_5 U_{m^{\prime}m}(-x)\right\} {\rm Tr}\!\left\{\gamma_\mu  S_{jj^{\prime}}(x)\gamma_\nu C S_{ii^{\prime}}^T(x)C \right\} \gamma_5 \gamma^\mu U_{kk^{\prime}}(x) \gamma^\nu\gamma_5\\
+{\rm Tr}\! \left\{i\gamma_5 C_{mm^{\prime}}(x)i\gamma_5 D_{m^{\prime}m}(-x)\right\} {\rm Tr}\!\left\{\gamma_\mu  S_{jj^{\prime}}(x)\gamma_\nu C S_{ii^{\prime}}^T(x)C \right\} \gamma_5 \gamma^\mu D_{kk^{\prime}}(x) \gamma^\nu\gamma_5
\end{array}\right\} \, ,\nonumber\\
\end{eqnarray}
\begin{eqnarray}
\Pi^{D^*N}_{\mu\nu}(p)&=&i\varepsilon^{ijk}\varepsilon^{i^{\prime}j^{\prime}k^{\prime}}\int d^4x e^{ip \cdot x}\nonumber\\
&&\left\{\begin{array}{l}+{\rm Tr}\! \left\{\gamma_\mu C_{mm^{\prime}}(x)\gamma_\nu U_{m^{\prime}m}(-x)\right\} {\rm Tr}\!\left\{\gamma_\alpha  U_{jj^{\prime}}(x)\gamma_\beta C U_{ii^{\prime}}^T(x)C \right\} \gamma_5 \gamma^\alpha D_{kk^{\prime}}(x) \gamma^\beta\gamma_5\\
+{\rm Tr}\! \left\{\gamma_\mu C_{mm^{\prime}}(x)\gamma_\nu D_{m^{\prime}m}(-x)\right\} {\rm Tr}\!\left\{\gamma_\alpha  D_{jj^{\prime}}(x)\gamma_\beta C D_{ii^{\prime}}^T(x)C \right\} \gamma_5 \gamma^\alpha U_{kk^{\prime}}(x) \gamma^\beta\gamma_5
\end{array}\right\} \, ,\nonumber\\
\end{eqnarray}
\begin{eqnarray}
\Pi^{D^*\Xi}_{\mu\nu}(p)&=&i\varepsilon^{ijk}\varepsilon^{i^{\prime}j^{\prime}k^{\prime}}\int d^4x e^{ip \cdot x}\nonumber\\
&&\left\{\begin{array}{l}+{\rm Tr}\! \left\{\gamma_\mu C_{mm^{\prime}}(x)\gamma_\nu U_{m^{\prime}m}(-x)\right\} {\rm Tr}\!\left\{\gamma_\alpha  S_{jj^{\prime}}(x)\gamma_\beta C S_{ii^{\prime}}^T(x)C \right\} \gamma_5 \gamma^\alpha U_{kk^{\prime}}(x) \gamma^\beta\gamma_5\\
+{\rm Tr}\! \left\{\gamma_\mu C_{mm^{\prime}}(x)\gamma_\nu D_{m^{\prime}m}(-x)\right\} {\rm Tr}\!\left\{\gamma_\alpha  S_{jj^{\prime}}(x)\gamma_\beta C S_{ii^{\prime}}^T(x)C \right\} \gamma_5 \gamma^\alpha D_{kk^{\prime}}(x) \gamma^\beta\gamma_5
\end{array}\right\} \, ,\nonumber\\
\end{eqnarray}
where the $U_{ij}(x)$, $D_{ij}(x)$, $S_{ij}$ and $C_{ij}(x)$ are the full $u$, $d$, $s$ and $c$ quark propagators, respectively,
\begin{eqnarray}
U/D_{ij}(x)&=& \frac{i\delta_{ij}\!\not\!{x}}{ 2\pi^2x^4}-\frac{\delta_{ij}\langle
\bar{q}q\rangle}{12} -\frac{\delta_{ij}x^2\langle \bar{q}g_s\sigma Gq\rangle}{192} -\frac{ig_sG^{a}_{\alpha\beta}t^a_{ij}(\!\not\!{x}
\sigma^{\alpha\beta}+\sigma^{\alpha\beta} \!\not\!{x})}{32\pi^2x^2}  \nonumber\\
&&  -\frac{1}{8}\langle\bar{q}_j\sigma^{\mu\nu}q_i \rangle \sigma_{\mu\nu}+\cdots \, ,
\end{eqnarray}
\begin{eqnarray}
S_{ij}(x)&=& \frac{i\delta_{ij}\!\not\!{x}}{ 2\pi^2x^4}
-\frac{\delta_{ij}m_s}{4\pi^2x^2}-\frac{\delta_{ij}\langle
\bar{s}s\rangle}{12} +\frac{i\delta_{ij}\!\not\!{x}m_s
\langle\bar{s}s\rangle}{48}-\frac{\delta_{ij}x^2\langle \bar{s}g_s\sigma Gs\rangle}{192}+\frac{i\delta_{ij}x^2\!\not\!{x} m_s\langle \bar{s}g_s\sigma
 Gs\rangle }{1152}\nonumber\\
&& -\frac{ig_s G^{a}_{\alpha\beta}t^a_{ij}(\!\not\!{x}
\sigma^{\alpha\beta}+\sigma^{\alpha\beta} \!\not\!{x})}{32\pi^2x^2} -\frac{\delta_{ij}x^4\langle \bar{s}s \rangle\langle g_s^2 GG\rangle}{27648}-\frac{1}{8}\langle\bar{s}_j\sigma^{\mu\nu}s_i \rangle \sigma_{\mu\nu} +\cdots \, ,
\end{eqnarray}
\begin{eqnarray}
C_{ij}(x)&=&\frac{i}{(2\pi)^4}\int d^4k e^{-ik \cdot x} \left\{
\frac{\delta_{ij}}{\!\not\!{k}-m_c}
-\frac{g_sG^n_{\alpha\beta}t^n_{ij}}{4}\frac{\sigma^{\alpha\beta}(\!\not\!{k}+m_c)+(\!\not\!{k}+m_c)
\sigma^{\alpha\beta}}{(k^2-m_c^2)^2}\right.\nonumber\\
&&\left. -\frac{g_s^2 (t^at^b)_{ij} G^a_{\alpha\beta}G^b_{\mu\nu}(f^{\alpha\beta\mu\nu}+f^{\alpha\mu\beta\nu}+f^{\alpha\mu\nu\beta}) }{4(k^2-m_c^2)^5}+\cdots\right\} \, ,\nonumber\\
f^{\alpha\beta\mu\nu}&=&(\!\not\!{k}+m_c)\gamma^\alpha(\!\not\!{k}+m_c)\gamma^\beta(\!\not\!{k}+m_c)\gamma^\mu(\!\not\!{k}+m_c)\gamma^\nu(\!\not\!{k}+m_c)\, ,
\end{eqnarray}
and  $t^n=\frac{\lambda^n}{2}$, the $\lambda^n$ is the Gell-Mann matrix \cite{QCDSR-Reinders85,Pascual-1984,WZG-HT-PRD}.
We accomplish the operator product expansion with the full light-quark propagators and full heavy-quark  propagator up to the vacuum condensates of dimension 13 in a consistent way \cite{WZG-HT-PRD}, and select the tensor structures  $\!\not\!{p}$, $1$; $\!\not\!{p} g_{\mu\nu}$, $g_{\mu\nu}$; $\!\not\!{p} \left(g_{\mu\alpha}g_{\nu\beta}+g_{\mu\beta}g_{\nu\alpha}\right)  $, $g_{\mu\alpha}g_{\nu\beta}+g_{\mu\beta}g_{\nu\alpha}$ in the correlation functions  $\Pi(p)$; $\Pi_{\mu\nu}(p)$; $\Pi_{\mu\nu\alpha\beta}(p)$ to investigate the pentaquark molecular states with the spin-parity $J^P=\frac{1}{2}^\mp$; $\frac{3}{2}^\mp$; $\frac{5}{2}^\mp$, respectively. Again we  obtain   the QCD spectral densities  through  dispersion relation,
\begin{eqnarray}
\frac{{\rm Im}\Pi_{j}(s)}{\pi}&=&\!\not\!{p}\, \rho^1_{j,QCD}(s)+\rho^0_{j,QCD}(s) \, ,
\end{eqnarray}
where $j=\frac{1}{2}$, $\frac{3}{2}$, $\frac{5}{2}$.

Now  we  suppose quark-hadron duality below the continuum thresholds  $s_0$, again we resort to the weight  function $\exp\left(-\frac{s}{T^2}\right)$ to suppress the excited states and continuum states to obtain   the  QCD sum rules:
\begin{eqnarray}\label{QCDSR}
2M_{-}{\lambda^{-}_{j}}^2\exp\left( -\frac{M_{-}^2}{T^2}\right)
&=& \int_{m_c^2}^{s_0}ds \left[\sqrt{s}\rho^1_{j,QCD}(s)+\rho^0_{j,QCD}(s)\right]\exp\left( -\frac{s}{T^2}\right)\, .
\end{eqnarray}

We differentiate   Eq.\eqref{QCDSR} with respect to  $\tau=\frac{1}{T^2}$, then eliminate the pole residues $\lambda^{-}_{j}$ with $j=\frac{1}{2}$, $\frac{3}{2}$, $\frac{5}{2}$ to obtain the QCD sum rules for the masses of the singly-charmed pentaquark molecular states,
\begin{eqnarray}
M^2_{-} &=& \frac{-\frac{d}{d \tau}\int_{m_c^2}^{s_0}ds \,\left[\sqrt{s}\,\rho^1_{QCD}(s)+\,\rho^0_{QCD}(s)\right]\exp\left(- \tau s\right)}{\int_{m_c^2}^{s_0}ds \left[\sqrt{s}\,\rho_{QCD}^1(s)+\,\rho^0_{QCD}(s)\right]\exp\left( -\tau s\right)}\, ,
\end{eqnarray}
where the spectral densities $\rho_{QCD}^1(s)=\rho_{j,QCD}^1(s)$ and $\rho^0_{QCD}(s)=\rho^0_{j,QCD}(s)$. As the expressions of the spectral densities are lengthy and the paper only has limited space, we only give the  analytic expressions for the $DN$, $D^*N$, $D\Xi$ and $D^*\Xi$ states in the Appendix, the other  analytic expressions can be obtained by emailing the corresponding author.

\section{Numerical results and discussions}
We take  the standard values of the  vacuum condensates
$\langle\bar{q}q \rangle=-(0.24\pm 0.01\, \rm{GeV})^3$,  $\langle\bar{s}s \rangle=(0.8\pm0.1)\langle\bar{q}q \rangle$,
 $\langle\bar{q}g_s\sigma G q \rangle=m_0^2\langle \bar{q}q \rangle$, $\langle\bar{s}g_s\sigma G s \rangle=m_0^2\langle \bar{s}s \rangle$,
$m_0^2=(0.8 \pm 0.1)\,\rm{GeV}^2$, $\langle \frac{\alpha_s
GG}{\pi}\rangle=0.012\pm0.004\,\rm{GeV}^4$    at the energy scale  $\mu=1\, \rm{GeV}$
\cite{QCDSR-SVZ79,QCDSR-Reinders85,QCDSR-Colangelo-Review}, and  take the $\overline{MS}$ masses $m_{c}(m_c)=(1.275\pm0.025)\,\rm{GeV}$ and $m_s(\mu=2\,\rm{GeV})=(0.095\pm0.005)\,\rm{GeV}$ from the Particle Data Group \cite{PDG-2020}. Furthermore,  we take account of the energy-scale dependence of  the quark condensates, mixed quark condensates and $\overline{MS}$ masses according to  the renormalization group equation \cite{Narison-mix},
\begin{eqnarray}
 \langle\bar{q}q \rangle(\mu)&=&\langle\bar{q}q\rangle({\rm 1 GeV})\left[\frac{\alpha_{s}({\rm 1 GeV})}{\alpha_{s}(\mu)}\right]^{\frac{12}{33-2n_f}}\, , \nonumber\\
 \langle\bar{s}s \rangle(\mu)&=&\langle\bar{s}s \rangle({\rm 1 GeV})\left[\frac{\alpha_{s}({\rm 1 GeV})}{\alpha_{s}(\mu)}\right]^{\frac{12}{33-2n_f}}\, , \nonumber\\
 \langle\bar{q}g_s \sigma Gq \rangle(\mu)&=&\langle\bar{q}g_s \sigma Gq \rangle({\rm 1 GeV})\left[\frac{\alpha_{s}({\rm 1 GeV})}{\alpha_{s}(\mu)}\right]^{\frac{2}{33-2n_f}}\, ,\nonumber\\
  \langle\bar{s}g_s \sigma Gs \rangle(\mu)&=&\langle\bar{s}g_s \sigma Gs \rangle({\rm 1 GeV})\left[\frac{\alpha_{s}({\rm 1 GeV})}{\alpha_{s}(\mu)}\right]^{\frac{2}{33-2n_f}}\, ,\nonumber\\
m_c(\mu)&=&m_c(m_c)\left[\frac{\alpha_{s}(\mu)}{\alpha_{s}(m_c)}\right]^{\frac{12}{33-2n_f}} \, ,\nonumber\\
m_s(\mu)&=&m_s({\rm 2GeV} )\left[\frac{\alpha_{s}(\mu)}{\alpha_{s}({\rm 2GeV})}\right]^{\frac{12}{33-2n_f}}\, ,\nonumber\\
\alpha_s(\mu)&=&\frac{1}{b_0t}\left[1-\frac{b_1}{b_0^2}\frac{\log t}{t} +\frac{b_1^2(\log^2{t}-\log{t}-1)+b_0b_2}{b_0^4t^2}\right]\, ,
\end{eqnarray}
where $t=\log \frac{\mu^2}{\Lambda^2}$, $b_0=\frac{33-2n_f}{12\pi}$, $b_1=\frac{153-19n_f}{24\pi^2}$, $b_2=\frac{2857-\frac{5033}{9}n_f+\frac{325}{27}n_f^2}{128\pi^3}$,  $\Lambda=213\,\rm{MeV}$, $296\,\rm{MeV}$  and  $339\,\rm{MeV}$ for the flavor numbers  $n_f=5$, $4$ and $3$, respectively  \cite{PDG-2020,Narison-mix}. We choose the quark flavor numbers $n_f = 4$ when exploring  the singly-charmed molecular states.

In order to choose the suitable energy scales, we take  the modified  energy scale formula $ \mu =\sqrt{M_{P}^2-{\mathbb{M}}_c^2}-k{\mathbb{M}}_s$, where $M_P=M_{-}$, the effective $c$-quark mass  ${\mathbb{M}}_c=1.82\,\rm{GeV}$ and effective $s$-quark mass  ${\mathbb{M}}_s=0.2\,\rm{GeV}$, the $k$ is the number of the $s$-quark in the currents/states, the effective masses are fitted in the QCD sum rules for the tetraquark  (molecular) states \cite{WZG-PRD-cc,WZG-IJMPA-mole}. In calculations, we adopt the rule, if the energy scale from the formula for a current containing three strange quarks is smaller than that of the corresponding  current containing two strange quarks, we opt for the larger energy scale (thus uniform energy scale for the cousin) to consistent with our naive expectation.

\begin{figure}
 \centering
  \includegraphics[totalheight=5cm,width=7cm]{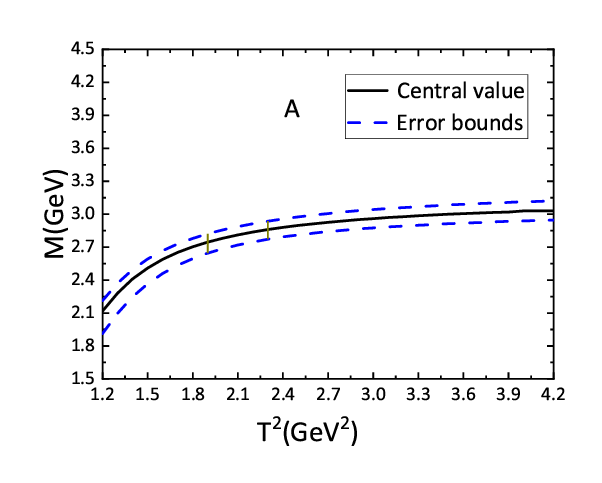}
 \includegraphics[totalheight=5cm,width=7cm]{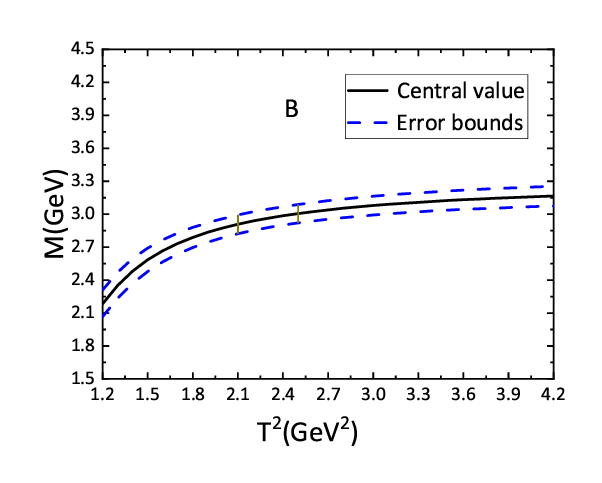}
   \includegraphics[totalheight=5cm,width=7cm]{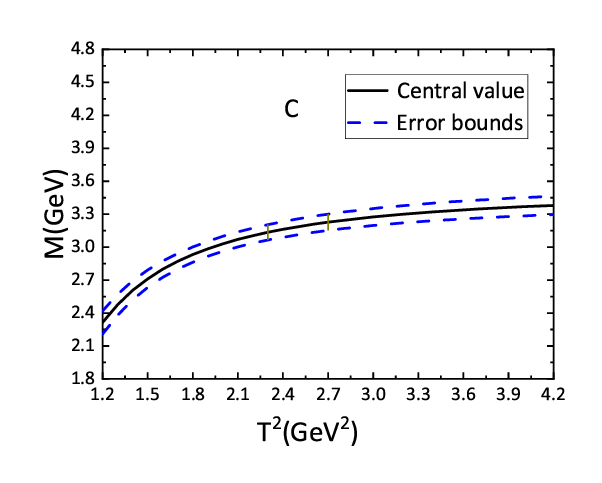}
 \includegraphics[totalheight=5cm,width=7cm]{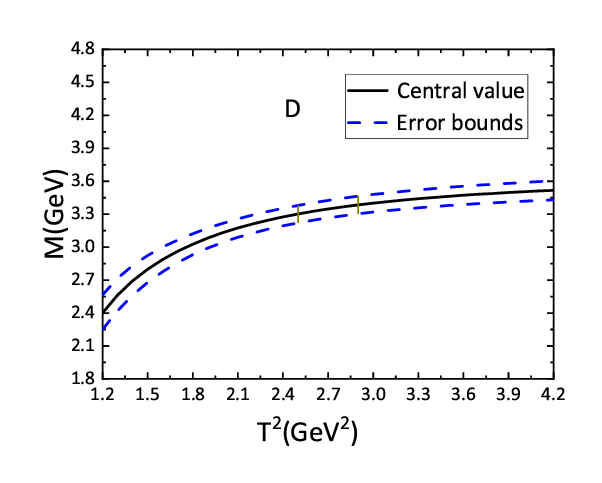}
 \caption{ The masses of the charmed  pentaquark molecular states with variations of the Borel parameters $T^2$, where the $A$, $B$, $C$ and $D$ denote the $DN$, $D^*N$, $D\Xi$ and $D^*\Xi$ molecular states, respectively.  }\label{Mass-Borel}
\end{figure}

 Now we define the pole contributions (PC) as
\begin{eqnarray}
{\rm PC}&=& \frac{  \int_{m_c^2}^{s_0} ds\,\left[\sqrt{s}\rho_{QCD}^1(s)+  \rho_{QCD}^0(s)\right]\,
\exp\left(-\frac{s}{T^2}\right)}{\int_{m_c^2}^{\infty} ds \,\left[\sqrt{s}\rho_{QCD}^1(s)+ \rho_{QCD}^0(s)\right]\,\exp\left(-\frac{s}{T^2}\right)}\, .
\end{eqnarray}
In calculations, we pick the central values of the $\rm{PC}$ at about $50\%$ in a systematic way and choose a uniform standard for the continuous threshold parameters $\sqrt{s_0}=M_{\rm{c}}+ (0.65-0.70)\,\rm{GeV}$, where the subscript $M_{\rm{c}}$ denotes the central values of the masses. Through continual trial and error, the energy scales $\mu$, the Borel parameters  $T^2$, the continuum threshold parameters $s_0$ and the pole contributions  are obtained and shown  in Table \ref{paramaters}. We choose the Borel windows in the range of $T^2_{max}-T^2_{min}=0.4\,\rm{GeV}^2$, and plot the masses versus Borel parameters, where the Borel windows  lie between the two vertical lines, see Fig.\ref{Mass-Borel}.

We can borrow some ideas from the experimental data from the Particle Data Group \cite{PDG-2020}, where the energy gaps between the ground states and first radial excited states for the $D_s$, $\eta_c$ and $J/\psi$ mesons are $0.62\,\rm{GeV}$, $0.65\,\rm{GeV}$ and $0.59\,\rm{GeV}$, respectively, the energy gaps between the ground states and first radial excited states for the $\Lambda_c$ and $\Xi_c$ baryons are about $0.50\,\rm{GeV}$. So the energy gaps $0.65-0.70\,\rm{GeV}$ are expected to be reasonable for the singly-charmed  pentaquark molecules, even if there are some contaminations from the higher resonances or continuum states, the contaminations are very small as $\exp(-\frac{s_0}{T^2})\ll 1\%$. In our previous works, we have proven that the contributions of the two-meson (two-baryon) scattering states can be safely absorbed into the pole residues and cannot affect the predicted masses remarkably for the tetraquark states (dibaryon states) \cite{WZG-IJMPA-Zc3900,WZG-PRD-Landau} (\cite{WZG-PRD-dibaryon}). We expect such conclusion survives in the present case and all the meson-baryon scattering states can be safely absorbed into the pole residues and cannot affect the predicted masses considerably.

\begin{figure}
 \centering
  \includegraphics[totalheight=5cm,width=7cm]{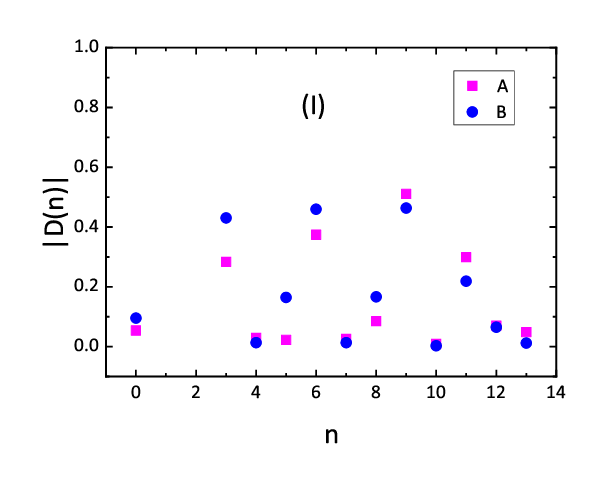}
  \includegraphics[totalheight=5cm,width=7cm]{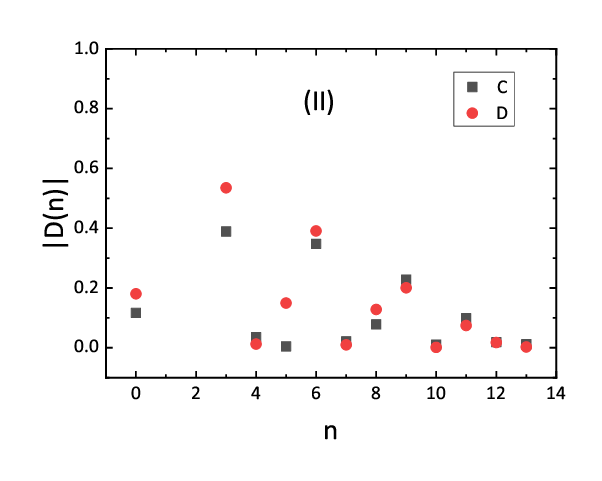}
 \caption{The absolute values of the contributions of the vacuum condensates of dimension $n$ for the central values of the  input  parameters for the charmed  pentaquark molecular states, where the $A$, $B$, $C$ and $D$ denote the  $DN$, $D^*N$, $D\Xi$  and $D^*\Xi$ states, respectively.}\label{dn}
\end{figure}

The contributions of different terms in the  operator product expansion are defined as,
\begin{eqnarray}
D(n)&=& \frac{  \int_{m_c^2}^{s_0} ds\,\rho_{n}(s)\,\exp\left(-\frac{s}{T^2}\right)}{\int_{m_c^2}^{s_0} ds \,\rho(s)\,\exp\left(-\frac{s}{T^2}\right)}\, ,
\end{eqnarray}
where the $\rho_{n}(s)$ are the QCD spectral densities for the vacuum condensates of dimension $n$, and the total spectral densities $\rho(s)=\sqrt{s}\rho^1_{QCD}(s)+ \rho^0_{QCD}(s)$. In Fig.\ref{dn}, we plot the absolute values of the contributions of the vacuum condensates  $|D(n)|$ for the $DN$, $D^*N$, $D\Xi$ and $D^*\Xi$ molecular states for the central values of the input parameters shown in Table \ref{paramaters}, as can be seen,  although the largest contributions come from the vacuum condensates of dimensions $n=3$, $6$, $9$,
the highest dimensional vacuum   condensate contributions have the relation $|D(11)|\gg|D(12)|>|D(13)|$ for all the singly-charmed pentaquark molecules,  the operator product expansion is convergent.

\begin{table}
\begin{center}
\begin{tabular}{|c|c|c|c|c|c|c|c|c|}\hline\hline
$$      &$J^P$  &\rm{Isospin}      &$\mu$                &$T^2 (\rm{GeV}^2)$   &$\sqrt{s_0}(\rm GeV) $    &pole          &$|D(13)|$

\\\hline

$DN$                 &$\frac{1}{2}^-$   &$0$            &$2.1$        & $1.9-2.3$            &$3.45\pm0.10$       &$(41-70)\%$       &$4.9\%$           \\ \hline

$DN$                 &$\frac{1}{2}^-$   &$1$            &$2.1$        & $1.9-2.3$            &$3.45\pm0.10$       &$(41-70)\%$       &$4.9\%$     \\ \hline

$D\Xi$               &$\frac{1}{2}^-$   &$0$            &$2.2$        &$2.3-2.7$           &$3.85\pm0.10$       &$(41-67)\%$       &$1.2\%$   \\ \hline

$D\Xi$               &$\frac{1}{2}^-$   &$1$            &$2.2$        &$2.3-2.7$          &$3.85\pm0.10$       &$(41-67)\%$       &$1.2\%$    \\ \hline

$D_s\Xi$             &$\frac{1}{2}^-$   &$\frac{1}{2}$  &$2.2$       &$2.4-2.8$             &$3.95\pm0.10$       &$(41-65)\%$       &$0.9\%$       \\ \hline

$D^*N$               &$\frac{3}{2}^-$   &$0$             &$2.3$      & $2.1-2.5$           &$3.60\pm0.10$       &$(37-64)\%$       &$1.2\%$         \\ \hline

$D^*N$               &$\frac{3}{2}^-$   &$1$             &$2.3$      & $2.1-2.5$           &$3.60\pm0.10$       &$(37-64)\%$       &$1.2\%$       \\ \hline

$D^*\Xi$             &$\frac{3}{2}^-$   &$0$             &$2.4$      & $2.5-2.9$            &$4.00\pm0.10$       &$(38-62)\%$       &$0.3\%$         \\ \hline

$D^*\Xi$             &$\frac{3}{2}^-$   &$1$              &$2.4$      & $2.5-2.9$            &$4.00\pm0.10$       &$(38-62)\%$       &$0.3\%$ \\ \hline

$D_s^*\Xi$           &$\frac{3}{2}^-$   &$\frac{1}{2}$    &$2.4$     & $2.5-2.9$            &$4.05\pm0.10$       &$(38-62)\%$       &$0.3\%$\\ \hline

$D\Xi^*$             &$\frac{3}{2}^-$   &$0$              &$2.5$     & $2.5-2.9$           &$4.10\pm0.10$       &$(41-66)\%$       &$0.5\%$     \\ \hline

$D\Xi^*$             &$\frac{3}{2}^-$   &$1$              &$2.5$     & $2.5-2.9$           &$4.10\pm0.10$       &$(41-66)\%$       &$0.5\%$  \\ \hline

$D_s\Xi^*$           &$\frac{3}{2}^-$   &$\frac{1}{2}$    &$2.5$     & $2.6-3.0$            &$4.20\pm0.10$       &$(40-64)\%$       &$0.5\%$ \\  \hline

$D^*\Xi^*$           &$\frac{5}{2}^-$   &$0$              &$2.6$     & $2.6-3.0$            &$4.20\pm0.10$       &$(41-64)\%$       &$0.4\%$ \\  \hline

$D^*\Xi^*$           &$\frac{5}{2}^-$   &$1$              &$2.6$     & $2.6-3.0$            &$4.20\pm0.10$       &$(41-64)\%$       &$0.4\%$ \\ \hline

$D_s^*\Xi^*$        &$\frac{5}{2}^-$    &$\frac{1}{2}$    &$2.6$     & $2.8-3.2$         &$4.30\pm0.10$      &$(40-63)\%$        &$0.2\%$                   \\
\hline\hline
\end{tabular}
\end{center}
\caption{ The spin-parity, isospin, energy scales $\mu$, Borel parameters $T^2$, continuum threshold parameters $s_0$, pole contributions and  vacuum condensates  $|D(13)|$ for the charmed  pentaquark molecular states. }\label{paramaters}
\end{table}

\begin{table}
\begin{center}
\begin{tabular}{|c|c|c|c|c|c|c|c|c|}\hline\hline
$$       &$J^P$  &\rm{Isospin}      &$M (\rm{GeV})$    &$\lambda (10^{-4}\rm{GeV}^6) $
&Thresholds\,(\rm{MeV}) &Assignments

\\ \hline

$DN$              &$\frac{1}{2}^-$           &$0$        &$2.81^{+0.13}_{-0.16}$   &$7.98^{+1.98}_{-1.56}$           &$2806$       &                \\ \hline

$DN$              &$\frac{1}{2}^-$           &$1$        &$2.81^{+0.13}_{-0.16}$   &$7.98^{+1.98}_{-1.56}$           &$2806$       & ? $\Sigma_c(2800)$           \\ \hline

$D\Xi$            &$\frac{1}{2}^-$           &$0$        & $3.19^{+0.11}_{-0.13}$           &$14.22^{+2.74}_{-2.40}$          &$3186$       & ? $\Omega_c(3185)$        \\ \hline

$D\Xi$            &$\frac{1}{2}^-$           &$1$        & $3.19^{+0.11}_{-0.13}$           &$14.22^{+2.74}_{-2.40}$          &$3186$       &                \\ \hline

$D_s\Xi$          &$\frac{1}{2}^-$      &$\frac{1}{2}$   &$3.28^{+0.12}_{-0.12}$           &$16.04^{+3.04}_{-2.66}$         &$3287$    &                          \\ \hline

$D^*N$            &$\frac{3}{2}^-$           &$0$        &$2.96^{+0.13}_{-0.14}$           &$8.87^{+1.84}_{-1.59}$          &$2947$     & ?? $\Lambda_c(2940)/\Lambda_c(2910)$
\\ \hline

$D^*N$            &$\frac{3}{2}^-$           &$1$         &$2.96^{+0.13}_{-0.14}$           &$8.87^{+1.84}_{-1.59}$          &$2947$     &                       \\ \hline

$D^*\Xi$          &$\frac{3}{2}^-$           &$0$         &$3.35^{+0.11}_{-0.13}$          &$15.89^{+2.93}_{-2.56}$         &$3327$     & ? $\Omega_c(3327)$    \\ \hline

$D^*\Xi$          &$\frac{3}{2}^-$           &$1$         & $3.35^{+0.11}_{-0.13}$          &$15.89^{+2.93}_{-2.56}$         &$3327$     &                                 \\ \hline

$D_s^*\Xi$        &$\frac{3}{2}^-$     &$\frac{1}{2}$     &$3.44^{+0.11}_{-0.11}$             &$17.24^{+3.06}_{-2.66}$
&$3430$    &          \\  \hline

$D\Xi^*$          &$\frac{3}{2}^-$          &$0$          & $3.41^{+0.11}_{-0.12}$
&$10.67^{+1.92}_{-1.68}$         &$3399$    &   \\ \hline

$D\Xi^*$          &$\frac{3}{2}^-$          &$1$          & $3.41^{+0.11}_{-0.12}$
&$10.67^{+1.92}_{-1.68}$         &$3399$       &   \\ \hline

$D_s\Xi^*$        &$\frac{3}{2}^-$     &$\frac{1}{2}$      & $3.51^{+0.11}_{-0.13}$
&$11.93^{+2.11}_{-1.85}$         &$3500$    &       \\  \hline

$D^*\Xi^*$        &$\frac{5}{2}^-$         &$0$             &$3.54^{+0.11}_{-0.12}$                &$10.93^{+1.94}_{-1.71}$         &$3540$    &                       \\  \hline

$D^*\Xi^*$        &$\frac{5}{2}^-$         &$1$             &$3.54^{+0.11}_{-0.12}$                &$10.93^{+1.94}_{-1.71}$         &$3540$             &              \\ \hline

$D_s^*\Xi^*$      &$\frac{5}{2}^-$      &$\frac{1}{2}$      & $3.65^{+0.11}_{-0.11}$        &$12.91^{+2.19}_{-1.93}$         &$3644$              &                       \\
\hline\hline
\end{tabular}
\end{center}
\caption{ The masses, pole residues, (corresponding meson-baryon) thresholds and possible assignments for the charmed pentaquark  molecular states. }\label{mass-residue}
\end{table}

The resulting masses and pole residues of the pentaquark molecular states after taking account of all the uncertainties of the input parameters  are shown in Table \ref{mass-residue}. From Tables \ref{paramaters}-\ref{mass-residue}, we can see clearly that the  modified  energy scale formula $\mu= \sqrt{M_{P}^2-{\mathbb{M}}_c^2}-k{\mathbb{M}}_s$ is satisfied in most cases,   while the energy scales $ \sqrt{M_{P}^2-{\mathbb{M}}_c^2}-3{\mathbb{M}}_s$ for the $D_s\Xi$, $D_s^*\Xi$ and $D_s\Xi^*$ states are smaller than the  energy scales $ \sqrt{M_{P}^2-{\mathbb{M}}_c^2}-2{\mathbb{M}}_s$ for the corresponding  $D\Xi$, $D^*\Xi$ and $D\Xi^*$ states, respectively,  we choose the same energy scales $ \sqrt{M_{P}^2-{\mathbb{M}}_c^2}-2{\mathbb{M}}_s$ for those cousins so as to meet our naive expectations, i.e. larger masses correspond to larger energy scales.

From Table \ref{mass-residue}, we can see explicitly that the pentaquark molecular states with the isospin $(I,I_3)=(1,0)$ and $(0,0)$ have the degenerated masses and pole residues. If we examine the currents in Eqs.\eqref{current-DN}-\eqref{current-DN-T}, we can see clearly that there exist symbolic quark structures $\frac{\bar{u}u-\bar{d}d}{\sqrt{2}}$ and $\frac{\bar{u}u+\bar{d}d}{\sqrt{2}}$ in the currents with the isospin $(I,I_3)=(1,0)$ and $(0,0)$, respectively, which leads to the same expressions for the QCD spectral densities, just like in the case of the QCD sum rules for the $\rho^0$ and $\omega$, and we have to resort to fine-tuning to adjust the hadron parameters so as to reproduce  the experimental data. However, at the present case, there lack experimental data, and we choose the same parameters and obtain degenerated masses and pole residues.

The predictions  $M=2.81^{+0.13}_{-0.16}~\rm{GeV}$ and $3.19^{+0.11}_{-0.13}~\rm{GeV}$ for the $DN$ and $D\Xi$ pentaquark molecular states with the $J^P =\frac{1}{2}^-$ support interpreting the $\Sigma_c(2800)$ and $\Omega_c(3185)$ as the $DN$ and $D\Xi$  molecule states, respectively. We also see that the mass of the $D^*\Xi$ molecular  state with the $J^P =\frac{3}{2}^-$ is $3.35^{+0.11}_{-0.13}~\rm{GeV}$, which is well consistent with the $\Omega_c(3327)$ observed  by  the LHCb collaboration, and supports  interpreting the $\Omega_c(3327)$  as the $D^*\Xi$ molecular  state. The predicted mass  $2.96^{+0.14}_{-0.13}~\rm{GeV}$ for the $D^*N$ molecular state is consistent with the experimentally measured $\Lambda_c(2940)/\Lambda_c(2910)$ within uncertainties, which does not exclude the possibility that they are molecular states.

In Ref.\cite{ZJR-siglam}, J. R. Zhang takes the resonances $\Sigma_{c}(2800)$ and $\Lambda_{c}(2940)$ to be the $S$-wave $DN$ and $D^{*}N$ molecule candidates, respectively, and  investigates them with the QCD sum rules. He obtains the masses $3.64\pm0.33~\mbox{GeV}$ and $3.73\pm0.35~\mbox{GeV}$
for the $S$-wave $DN$ state with the $J^{P}=\frac{1}{2}^{-}$
 and  $S$-wave $D^{*}N$ state with the  $J^{P}=\frac{3}{2}^{-}$, respectively, which  are somewhat bigger than the experimental data of the $\Sigma_{c}(2800)$ and $\Lambda_{c}(2940)$,
respectively, and differ from the present calculations significantly.   J. R. Zhang takes the vacuum condensates and $c$-quark mass at the energy scales $\mu=1\,\rm{GeV}$ and   $m_c(m_c)$, respectively. While in this work, we determine the energy scales of the QCD spectral densities in a consistent way using the
 modified  energy scale formula $\mu= \sqrt{M_{P}^2-{\mathbb{M}}_c^2}-k{\mathbb{M}}_s$, which can enhance the pole contributions remarkably and improve the convergent behaviors  of the operator product expansion remarkably,  thus lead to smaller ground state masses. It is the main reason for the different predictions.

Taking the $D\Xi$ ($D^*\Xi$)  pentaquark molecular state with the $J^P={\frac{1}{2}}^-$ (${\frac{3}{2}}^-$)   as an example, from the angular momentum addition rules,  we know that the combination of the isospin  can lead to a total isospin of $1$ or $0$,  we obtain a symmetric
isotriplet  $| 1,1 \,\rangle$, $|1,0 \rangle$, $|1,-1\,\rangle$ and an antisymmetric isosinglet $| 0,0 \,\rangle$. For the $D\Xi$ ($D^*\Xi$) molecular state with the isospin $| 0,0\, \rangle$, we suppose it is a good candidate for the   $\Omega_c(3185)$ ($\Omega_c(3327)$).  If the corresponding molecular states with the isospin $| 1,1 \,\rangle$ and $| 1,-1 \,\rangle$ are observed in  the $\Xi_c^+\bar{K}^0$  and $\Xi_c^0\bar{K}^-$ mass spectrum respectively in the future,  it will prove that the $D\Xi$ ($D^*\Xi$) molecular state with the isospin $| 1,0 \,\rangle$  exists indeed and has the mass about  $3.19^{+0.11}_{-0.13}~\rm{GeV}$ ($3.35^{+0.11}_{-0.13}~\rm{GeV}$), thus shed light on the nature of the $\Omega_c(3185)$ ($\Omega_c(3327)$).

Other interpretations also exist, the $\Sigma_c(2800)$ is interpreted as overlap of the resonance  states $\Sigma_c (2813)$ and $\Sigma_c(2840)$ \cite{ZXH-2800-1} or P-wave charmed baryon state with the   $J^P=\frac{1}{2}^- /\frac{3}{2}^-/\frac{5}{2}^-$ \cite{ZXH-2800-2,CHX-2800,JDJ-2800}. Cheng et al  believe that the $\Sigma_c(2800)$  is not possible to be $J^P=\frac{1}{2}^-$ charmed baryon state \cite{CHY-2800-2940}.
In addition, the $\Lambda_c(2940)^+$ is most likely to be the $J^P=\frac{1}{2}^-$ or $\frac{3}{2}^-$ (2P) state \cite{CHY-2800-2940,LX-2940,LQF-2940-1}.
Moreover, the $\Lambda_c(2910)^+$ is probably the $J^P=\frac{1}{2}^-$ (2P) \cite{Azizi-2910} or $\frac{5}{2}^-$ (1P) state \cite{ZXH-2910}.
Furthermore, the $\Omega_c(3185)$ lies in the region of the 2S state \cite{YGL-baryonQ,Karliner-3185-3327}, while the $\Omega_c(3327)$ is a good candidate for the 2S $\Omega_c$ state \cite{Karliner-3185-3327} or D-wave  $\Omega_c$ baryon state with the $J^P=\frac{1}{2}^+/\frac{3}{2}^+/\frac{5}{2}^+$ \cite{YGL-baryonQ,LX-3327,WZG-3327}.

\section{Conclusion}
Inspired by observation of  the new baryon states $\Omega_c(3185)$ and $\Omega_c(3327)$ by the LHCb collaboration, we use the QCD sum rules to study a series of singly-charmed  molecular states consisting of the color-neutral clusters having the same quantum numbers as the mesons ($D$, $D^*$, $D_s$ and $D_s^*$) and baryons ($N$, $\Xi$ and $\Xi^*$). In calculations, we distinguish the isospin explicitly, and distinguish the contributions from the negative-parity and positive-parity molecular states explicitly, it is a unique feature of our work.       The numerical results favor assigning the $\Omega_c(3185)$ as the $D\Xi$ molecular state  with the $J^P=\frac{1}{2}^-$ and $| I,I_3 \rangle=| 0,0 \rangle$,  assigning the $\Omega_c(3327)$ as the $D^*\Xi$ molecular state with the $J^P=\frac{3}{2}^-$ and $|I,I_3 \rangle=| 0,0 \rangle$,   assigning the $\Sigma_c(2800)$   as the $DN$ molecular state with the $J^P=\frac{1}{2}^-$ and $| I,I_3 \rangle=| 1,0 \rangle$,  and assigning the $\Lambda_c(2940)/\Lambda_c(2910)$ as the $D^*N$ molecular state with the $J^P=\frac{3}{2}^-$ and $| I,I_3 \rangle=| 0,0 \rangle$.  At the same time, we systematically predict the mass spectrum of the $D_s\Xi$, $D^*_s\Xi$, $D\Xi^*$, $D_s\Xi^*$, $D^*\Xi^*$ and $D_s^*\Xi^*$  pentaquark molecular states considering the light-flavor  $SU(3)$ breaking effects. We believe that the experimental collaborations could detect more singly-charmed baryon states which  could be candidates for those pentaquark molecular states. For example, we can search for the $D\Xi$ and $D^*\Xi$ molecular states with the isospin $| 1,\pm1 \,\rangle$  in  the $\Xi_c^+\bar{K}^0$  and $\Xi_c^0\bar{K}^-$ mass spectrum respectively in the future, which could shed light on the nature of the $\Omega_c(3185/3327)$.

\section*{Appendix}
The QCD spectral densities $\rho_{QCD}^1(s)$ and $\rho_{QCD}^0(s)$ for the $DN$, $D^*N$, $D\Xi$ and $D^*\Xi$ states,
\begin{eqnarray}
\rho_{QCD}^1(s)&=&\rho_{D\Xi}^1(s)\, ,\,\, \rho_{DN}^1(s)\, , \, \,\rho_{D^*\Xi}^1(s)\, , \, \,\rho_{D^*N}^0(s)\, , \nonumber \\
\rho_{QCD}^0(s)&=&\rho_{D\Xi}^0(s)\, ,\,\, \rho_{DN}^0(s)\, , \, \,\rho_{D^*\Xi}^0(s)\, , \, \,\rho_{D^*N}^1(s)\, .
\end{eqnarray}

{\bf  For the $D\Xi$ pentaquark molecular states},

\begin{eqnarray}
\rho^1_{D\Xi}(0)&=&\frac{1} {2457600\pi^8}\int_{x_i}^{1}dxx(1-x)^6(8s-3\tilde{m}_c^2)(s-\tilde{m}_c^2)^4\, ,
\end{eqnarray}

\begin{eqnarray}
\rho^1_{D\Xi}(3)&=&-\frac{m_c\langle\bar{q}q\rangle}{3072\pi^6}\int_{x_i}^{1}dx(1-x)^4(s-\tilde{m}_c^2)^3\, ,
\end{eqnarray}

\begin{eqnarray}
\rho^1_{D\Xi}(4)&=&\frac{m_c^2} {737280\pi^6}\langle\frac{\alpha_{s}GG}{\pi}\rangle\int_{x_i}^{1}dx\frac{(1-x)^6}{x^2}(-5s+3\tilde{m}_c^2)(s-\tilde{m}_c^2)\nonumber\\
&&+\frac{1} {81920\pi^6}\langle\frac{\alpha_{s}GG}{\pi}\rangle\int_{x_i}^{1}dx(1-x)^4(2+3x)(2s-\tilde{m}_c^2)(s-\tilde{m}_c^2)^2\, ,
\end{eqnarray}

\begin{eqnarray}
\rho^1_{D\Xi}(5)&=&\frac{m_s\langle\bar{s}g_s\sigma Gs\rangle}{6144\pi^6} \int_{x_i}^{1}dxx(1-x)^3(-5s+3\tilde{m}_c^2)(s-\tilde{m}_c^2)\nonumber\\
&&-\frac{m_c\langle\bar{q}g_s\sigma Gq\rangle}{2048\pi^6}\int_{x_i}^{1}dx\frac{(3x-1)(x-1)^3}{x}(s-\tilde{m}_c^2)^2\, ,
\end{eqnarray}

\begin{eqnarray}
\rho^1_{D\Xi}(6)&=&\frac{\langle\bar{s}s\rangle^2}{384\pi^4}\int_{x_i}^{1}dxx(1-x)^3(5s-3\tilde{m}_c^2)(s-\tilde{m}_c^2)\, ,
\end{eqnarray}

\begin{eqnarray}
\rho^1_{D\Xi}(7)&=&\frac{m_c^3\langle\bar{q}q\rangle}{9216\pi^4}\langle\frac{\alpha_{s}GG}{\pi}\rangle\int_{x_i}^{1}dx\frac{(1-x)^4}{x^3}\nonumber\\
&&-\frac{m_c\langle\bar{q}q\rangle}{3072\pi^4}\langle\frac{\alpha_{s}GG}{\pi}\rangle\int_{x_i}^{1}dx\frac{(1-x)^2(1-2x+9x^2)}{x^2}(s-\tilde{m}_c^2)\nonumber\\
&&+\frac{m_s\langle\bar{s}s\rangle}{15360\pi^4}\langle\frac{\alpha_{s}GG}{\pi}\rangle\int_{x_i}^{1}dxx(1-x)^2(551s-546\tilde{m}_c^2)\, ,
\end{eqnarray}

\begin{eqnarray}
\rho^1_{D\Xi}(8)&=&\frac{\langle\bar{s}s\rangle\langle\bar{s}g_s\sigma Gs\rangle}{256\pi^4}\int_{x_i}^{1}dxx(1-x)^2(-4s+3\tilde{m}_c^2)\nonumber\\
&&+\frac{m_sm_c\langle\bar{q}q\rangle\langle\bar{s}g_s\sigma Gs\rangle}{384\pi^4}\int_{x_i}^{1}dx(1-x)\, ,
\end{eqnarray}

\begin{eqnarray}
\rho^1_{D\Xi}(9)&=&-\frac{m_c\langle\bar{q}q\rangle\langle\bar{s}s\rangle^2}{24\pi^2}\int_{x_i}^{1}dx(1-x)\, ,
\end{eqnarray}

\begin{eqnarray}
\rho^1_{D\Xi}(10)&=&\frac{m_sm_c\langle\bar{q}g_s\sigma Gq\rangle\langle\bar{s}g_s\sigma Gs\rangle}{12288\pi^4}\int_{x_i}^{1}dx\frac{(17-25x)}{x}\delta(s-\tilde{m}_c^2)\nonumber\\
&&-\frac{\langle\bar{s}g_s\sigma Gs\rangle^2}{73728\pi^4}\int_{x_i}^{1}dx(1-x)^2s\delta(s-\tilde{m}_c^2)-\frac{\langle\bar{s}g_s\sigma Gs\rangle^2}{73728\pi^4}\int_{x_i}^{1}dx(x-1)(3x-7)\nonumber\\
&&-\frac{m_sm_c\langle\bar{q}q\rangle\langle\bar{s}s\rangle}{1152\pi^2}\langle\frac{\alpha_{s}GG}{\pi}\rangle\int_{x_i}^{1}dx\delta(s-\tilde{m}_c^2)\nonumber\\
&&+\frac{\langle\bar{s}s\rangle^2}{384\pi^2}\langle\frac{\alpha_{s}GG}{\pi}\rangle\int_{x_i}^{1}dx(x-3)(x-1)\nonumber\\
&&+\frac{\langle\bar{s}s\rangle^2}{1152\pi^2}\langle\frac{\alpha_{s}GG}{\pi}\rangle\int_{x_i}^{1}dx(x-3)(x-1)s\delta(s-\tilde{m}_c^2)\nonumber\\
&&-\frac{m_c^2\langle\bar{s}s\rangle^2}{1728\pi^2}\langle\frac{\alpha_{s}GG}{\pi}\rangle\int_{x_i}^{1}dx\frac{(1-x)^3}{x^2}\left(1+\frac{s}{2T^2}\right)\delta(s-\tilde{m}_c^2)\, ,
\end{eqnarray}

\begin{eqnarray}
\rho^1_{D\Xi}(11)&=&\frac{m_c\langle\bar{q}q\rangle\langle\bar{s}s\rangle\langle\bar{s}g_s\sigma Gs\rangle}{96\pi^2}\int_{x_i}^{1}dx\delta(s-\tilde{m}_c^2)\nonumber\\
&&+\frac{m_c\langle\bar{s}s\rangle^2\langle\bar{q}g_s\sigma Gq\rangle}{96\pi^2}\int_{x_i}^{1}dx\frac{3x-2}{x}\delta(s-\tilde{m}_c^2)\, ,
\end{eqnarray}

\begin{eqnarray}
\rho^1_{D\Xi}(13)&=&\frac{47m_c\langle\bar{s}s\rangle\langle\bar{q}g_s\sigma Gq\rangle\langle\bar{s}g_s\sigma Gs\rangle}{9216\pi^2T^2}\int_{x_i}^{1}dx\frac{1}{x}\delta(s-\tilde{m}_c^2)\nonumber\\
&&+\frac{m_c^3\langle\bar{q}q\rangle\langle\bar{s}s\rangle^2}{432T^4}\langle\frac{\alpha_{s}GG}{\pi}\rangle\int_{x_i}^{1}dx\frac{1-x}{x^3}\delta(s-\tilde{m}_c^2)\nonumber\\
&&-\frac{m_c\langle\bar{q}q\rangle\langle\bar{s}s\rangle^2}{144T^2}\langle\frac{\alpha_{s}GG}{\pi}\rangle\int_{x_i}^{1}dx\frac{1-x}{x^2}\delta(s-\tilde{m}_c^2)\nonumber\\
&&-\frac{m_c\langle\bar{q}q\rangle\langle\bar{s}s\rangle^2}{288T^2}\langle\frac{\alpha_{s}GG}{\pi}\rangle\delta(s-m_c^2)
\nonumber\\
&&-\frac{m_c\langle\bar{s}s\rangle\langle\bar{q}g_s\sigma Gq\rangle\langle\bar{s}g_s\sigma Gs\rangle}{384\pi^2T^2}\delta(s-m_c^2)\, ,
\end{eqnarray}

\begin{eqnarray}
\rho^0_{D\Xi}(3)&=&\frac{\langle\bar{q}q\rangle}{12288\pi^6}\int_{x_i}^{1}dxx(1-x)^4(-3s+\tilde{m}_c^2)(s-\tilde{m}_c^2)^3\, ,
\end{eqnarray}

\begin{eqnarray}
\rho^0_{D\Xi}(6)&=&\frac{m_c\langle\bar{q}q\rangle^2}{128\pi^4}\int_{x_i}^{1}dx(1-x)^2(s-\tilde{m}_c^2)^2\nonumber\\
&&+\frac{3m_s\langle\bar{q}q\rangle\langle\bar{s}s\rangle}{64\pi^4}\int_{x_i}^{1}dxx(1-x)^2(2s-\tilde{m}_c^2)(s-\tilde{m}_c^2)\, ,
\end{eqnarray}

\begin{eqnarray}
\rho^0_{D\Xi}(7)&=&\frac{m_c^2\langle\bar{q}q\rangle}{18432\pi^4}\langle\frac{\alpha_{s}GG}{\pi}\rangle\int_{x_i}^{1}dx\frac{(1-x)^4}{x^2}(3s-2\tilde{m}_c^2)\nonumber\\
&&+\frac{\langle\bar{q}q\rangle}{1536\pi^4}\langle\frac{\alpha_{s}GG}{\pi}\rangle\int_{x_i}^{1}dx(1-x)^2(2x-1)(2s-\tilde{m}_c^2)(s-\tilde{m}_c^2)\, ,
\end{eqnarray}

\begin{eqnarray}
\rho^0_{D\Xi}(8)&=&\frac{10m_s\langle\bar{q}q\rangle\langle\bar{s}g_s\sigma Gs\rangle}{384\pi^4}\int_{x_i}^{1}dxx(1-x)(-3s+2\tilde{m}_c^2)\nonumber\\
&&-\frac{7m_s\langle\bar{s}s\rangle\langle\bar{q}g_s\sigma Gq\rangle}{256\pi^4}\int_{x_i}^{1}dxx(1-x)(3s-2\tilde{m}_c^2)\nonumber\\
&&+\frac{m_c\langle\bar{q}q\rangle\langle\bar{q}g_s\sigma Gq\rangle}{128\pi^4}\int_{x_i}^{1}dx\frac{(x-1)(2x-1)}{x}(s-\tilde{m}_c^2)\, ,
\end{eqnarray}

\begin{eqnarray}
\rho^0_{D\Xi}(9)&=&\frac{\langle\bar{q}q\rangle\langle\bar{s}s\rangle^2}{12\pi^2}\int_{x_i}^{1}dxx(1-x)(-3s+2\tilde{m}_c^2)-\frac{m_sm_c\langle\bar{q}q\rangle^2\langle\bar{s}s\rangle}{8\pi^2}\int_{x_i}^{1}dx\, ,
\end{eqnarray}

\begin{eqnarray}
\rho^0_{D\Xi}(10)&=&\frac{m_c\langle\bar{q}g_s\sigma Gq\rangle^2}{12288\pi^4}\int_{x_i}^{1}dx\frac{1-x}{x}+\frac{m_s\langle\bar{q}g_s\sigma Gq\rangle\langle\bar{s}g_s\sigma Gs\rangle}{3072\pi^4}\int_{x_i}^{1}dx(1+43x)\nonumber\\
&&+\frac{m_s\langle\bar{q}g_s\sigma Gq\rangle\langle\bar{s}g_s\sigma Gs\rangle}{6144\pi^4}\int_{x_i}^{1}dxs(1+43x)\delta(s-\tilde{m}_c^2)\nonumber\\
&&-\frac{m_sm_c^2\langle\bar{s}s\rangle\langle\bar{q}q\rangle}{384\pi^2}\langle\frac{\alpha_{s}GG}{\pi}\rangle\int_{x_i}^{1}dx\frac{(1-x)^2}{x^2}\left(1+\frac{s}{T^2}\right)\delta(s-\tilde{m}_c^2)\nonumber\\
&&-\frac{m_c^3\langle\bar{q}q\rangle^2}{1152\pi^2}\langle\frac{\alpha_{s}GG}{\pi}\rangle\int_{x_i}^{1}dx\frac{(1-x)^2}{x^3}\delta(s-\tilde{m}_c^2)
\nonumber\\
&&+\frac{m_c\langle\bar{q}q\rangle^2}{1152\pi^2}\langle\frac{\alpha_{s}GG}{\pi}\rangle\int_{x_i}^{1}dx\frac{3-6x+2x^2}{x^2}\nonumber\\
&&+\frac{m_s\langle\bar{s}s\rangle\langle\bar{q}q\rangle}{576\pi^2}\langle\frac{\alpha_{s}GG}{\pi}\rangle\int_{x_i}^{1}dx(18-11x)\nonumber\\
&&+\frac{m_s\langle\bar{s}s\rangle\langle\bar{q}q\rangle}{1152\pi^2}\langle\frac{\alpha_{s}GG}{\pi}\rangle\int_{x_i}^{1}dxs(18-11x)\delta(s-\tilde{m}_c^2)\, ,
\end{eqnarray}

\begin{eqnarray}
\rho^0_{D\Xi}(11)&=&\frac{(2\langle\bar{q}q\rangle\langle\bar{s}s\rangle\langle\bar{s}g_s\sigma Gs\rangle+\langle\bar{q}g_s\sigma Gq\rangle\langle\bar{s}s\rangle^2)}{48\pi^2}\int_{x_i}^{1}dxx[2+s\delta(s-\tilde{m}_c^2)]\nonumber\\
&&+\frac{5m_sm_c\langle\bar{q}q\rangle^2\langle\bar{s}g_s\sigma Gs\rangle}{144\pi^2}\delta(s-m_c^2)+\frac{13m_sm_c\langle\bar{q}q\rangle\langle\bar{s}s\rangle\langle\bar{q}g_s\sigma Gq\rangle}{192\pi^2}\delta(s-m_c^2)\nonumber\\
&&-\frac{m_sm_c\langle\bar{q}q\rangle\langle\bar{s}s\rangle\langle\bar{q}g_s\sigma Gq\rangle}{16\pi^2}\int_{x_i}^{1}dx\frac{1}{x}\delta(s-\tilde{m}_c^2)\, ,
\end{eqnarray}

\begin{eqnarray}
\rho^0_{D\Xi}(12)&=&\frac{m_c\langle\bar{q}q\rangle^2\langle\bar{s}s\rangle^2}{9}\delta(s-m_c^2)\, ,
\end{eqnarray}

\begin{eqnarray}
\rho^0_{D\Xi}(13)&=&-\frac{\langle\bar{s}s\rangle\langle\bar{q}g_s\sigma Gq\rangle\langle\bar{s}g_s\sigma Gs\rangle}{4608\pi^2}\int_{x_i}^{1}dx\left(1+\frac{s}{T^2}\right)\delta(s-\tilde{m}_c^2)\nonumber\\
&&-\frac{(2\langle\bar{s}s\rangle\langle\bar{s}g_s\sigma Gs\rangle\langle\bar{q}g_s\sigma Gq\rangle+\langle\bar{s}g_s\sigma Gs\rangle^2\langle\bar{q}q\rangle)}{192\pi^2}\left(1+\frac{s}{T^2}\right)\delta(s-m_c^2)\nonumber\\
&&+\frac{m_sm_c\langle\bar{q}q\rangle\langle\bar{q}g_s\sigma Gq\rangle\langle\bar{s}g_s\sigma Gs\rangle}{1152\pi^2T^2}\left(20-\frac{21s}{T^2}\right)\delta(s-m_c^2)\nonumber\\
&&+\frac{m_sm_c\langle\bar{s}s\rangle\langle\bar{q}g_s\sigma Gq\rangle^2}{18432\pi^2T^2}\left(49+\frac{120s}{T^2}\right)\delta(s-m_c^2)\nonumber\\
&&-\frac{\langle\bar{q}q\rangle\langle\bar{s}s\rangle^2}{72}\langle\frac{\alpha_{s}GG}{\pi}\rangle\int_{x_i}^{1}dx\left(1+\frac{s}{T^2}\right)\delta(s-\tilde{m}_c^2)\nonumber\\
&&+\frac{m_sm_c^3\langle\bar{q}q\rangle^2\langle\bar{s}s\rangle}{144T^4}\langle\frac{\alpha_{s}GG}{\pi}\rangle\int_{x_i}^{1}dx\frac{1}{x^3}\delta(s-\tilde{m}_c^2)\nonumber\\
&&-\frac{m_sm_c\langle\bar{q}q\rangle^2\langle\bar{s}s\rangle}{48T^2}\langle\frac{\alpha_{s}GG}{\pi}\rangle\int_{x_i}^{1}dx\frac{1}{x^2}\delta(s-\tilde{m}_c^2)\nonumber\\
&&+\frac{m_c^2\langle\bar{q}q\rangle\langle\bar{s}s\rangle^2}{216T^4}\langle\frac{\alpha_{s}GG}{\pi}\rangle\int_{x_i}^{1}dx\frac{s(1-x)}{x^2}\delta(s-\tilde{m}_c^2)\nonumber\\
&&-\frac{5sm_sm_c\langle\bar{q}q\rangle^2\langle\bar{s}s\rangle}{432T^4}\langle\frac{\alpha_{s}GG}{\pi}\rangle\delta(s-m_c^2)\nonumber\\
&&-\frac{\langle\bar{q}q\rangle\langle\bar{s}s\rangle^2}{144}\langle\frac{\alpha_{s}GG}{\pi}\rangle\left(1+\frac{s}{T^2}\right)\delta(s-m_c^2)\, ,
\end{eqnarray}
where $\tilde{m}_c^2=\frac{{m}_c^2}{x}$, $x_i=\frac{{m}_c^2}{s}$.

{\bf  For the $DN$ pentaquark molecular states},
\begin{eqnarray}
\rho_{DN}^1(s)&=&\rho_{D\Xi}^1(s)\mid_{m_s \to 0, \langle\bar{s}s\rangle \to\langle\bar{q}q\rangle, \langle\bar{s}g_s\sigma Gs\rangle \to\langle\bar{q}g_s\sigma Gq\rangle} \, , \nonumber\\
\rho_{DN}^0(s)&=&\rho_{D\Xi}^0(s)\mid_{m_s \to 0, \langle\bar{s}s\rangle \to\langle\bar{q}q\rangle, \langle\bar{s}g_s\sigma Gs\rangle \to\langle\bar{q}g_s\sigma Gq\rangle} \, .
\end{eqnarray}

{\bf  For the $D^*\Xi$ pentaquark molecular states},
\begin{eqnarray}
\rho^1_{D^*\Xi}(0)&=&\frac{1} {2457600\pi^8}\int_{x_i}^{1}dxx(1-x)^6(-7s+2\tilde{m}_c^2)(s-\tilde{m}_c^2)^4\, ,
\end{eqnarray}

\begin{eqnarray}
\rho^1_{D^*\Xi}(3)&=&\frac{m_c\langle\bar{q}q\rangle}{3072\pi^6}\int_{x_i}^{1}dx(1-x)^4(s-\tilde{m}_c^2)^3\, ,
\end{eqnarray}

\begin{eqnarray}
\rho^1_{D^*\Xi}(4)&=&\frac{m_c^2} {368640\pi^6}\langle\frac{\alpha_{s}GG}{\pi}\rangle\int_{x_i}^{1}dx\frac{(1-x)^6}{x^2}(2s-\tilde{m}_c^2)(s-\tilde{m}_c^2)\nonumber\\
&&+\frac{1} {737280\pi^6}\langle\frac{\alpha_{s}GG}{\pi}\rangle\int_{x_i}^{1}dx(1-x)^4[s(14-89x)+(38x-8)\tilde{m}_c^2](s-\tilde{m}_c^2)^2\, ,
\end{eqnarray}

\begin{eqnarray}
\rho^1_{D^*\Xi}(5)&=&\frac{m_s\langle\bar{s}g_s\sigma Gs\rangle}{3072\pi^6} \int_{x_i}^{1}dxx(1-x)^3(2s-\tilde{m}_c^2)(s-\tilde{m}_c^2)\nonumber\\
&&-\frac{m_c\langle\bar{q}g_s\sigma Gq\rangle}{1024\pi^6}\int_{x_i}^{1}dx(1-x)^3(s-\tilde{m}_c^2)^2\, ,
\end{eqnarray}

\begin{eqnarray}
\rho^1_{D^*\Xi}(6)&=&-\frac{\langle\bar{s}s\rangle^2}{192\pi^4}\int_{x_i}^{1}dxx(1-x)^3(2s-\tilde{m}_c^2)(s-\tilde{m}_c^2)\, ,
\end{eqnarray}

\begin{eqnarray}	
\rho^1_{D^*\Xi}(7)&=&\frac{m_c\langle\bar{q}q\rangle}{3072\pi^4}\langle\frac{\alpha_{s}GG}{\pi}\rangle\int_{x_i}^{1}dx\frac{(1-x)^2(1-2x+9x^2)}{x^2}(s-\tilde{m}_c^2)\nonumber\\
&&-\frac{m_c^3\langle\bar{q}q\rangle}{9216\pi^4}\langle\frac{\alpha_{s}GG}{\pi}\rangle\int_{x_i}^{1}dx\frac{(1-x)^4}{x^3}\nonumber\\
&&-\frac{m_s\langle\bar{s}s\rangle}{3072\pi^4}\langle\frac{\alpha_{s}GG}{\pi}\rangle\int_{x_i}^{1}dxx(1-x)^2(3s-2\tilde{m}_c^2)\, ,
\end{eqnarray}

\begin{eqnarray}
\rho^1_{D^*\Xi}(8)&=&\frac{\langle\bar{s}s\rangle\langle\bar{s}g_s\sigma Gs\rangle}{256\pi^4}\int_{x_i}^{1}dxx(1-x)^2(3s-2\tilde{m}_c^2)\nonumber\\
&&-\frac{m_sm_c\langle\bar{q}q\rangle\langle\bar{s}g_s\sigma Gs\rangle}{384\pi^4}\int_{x_i}^{1}dx(1-x)\, ,
\end{eqnarray}

\begin{eqnarray}
\rho^1_{D^*\Xi}(9)&=&\frac{m_c\langle\bar{q}q\rangle\langle\bar{s}s\rangle^2}{24\pi^2}\int_{x_i}^{1}dx(1-x)\, ,
\end{eqnarray}

\begin{eqnarray}
\rho^1_{D^*\Xi}(10)&=&\frac{m_sm_c\langle\bar{q}q\rangle\langle\bar{s}s\rangle}{1152\pi^2}\langle\frac{\alpha_{s}GG}{\pi}\rangle\int_{x_i}^{1}dx\delta(s-\tilde{m}_c^2)\nonumber\\
&&+\frac{\langle\bar{s}s\rangle^2}{288\pi^2}\langle\frac{\alpha_{s}GG}{\pi}\rangle\int_{x_i}^{1}dx(1-x)(1-2x)\nonumber\\
&&+\frac{\langle\bar{s}s\rangle^2}{1152\pi^2}\langle\frac{\alpha_{s}GG}{\pi}\rangle\int_{x_i}^{1}dxs(1-x)(1-3x)\delta(s-\tilde{m}_c^2)\nonumber\\
&&+\frac{m_c^2\langle\bar{s}s\rangle^2}{3456\pi^2}\langle\frac{\alpha_{s}GG}{\pi}\rangle\int_{x_i}^{1}dx\frac{(1-x)^3}{x^2}\left(1+\frac{s}{T^2}\right)\delta(s-\tilde{m}_c^2)\nonumber\\
&&-\frac{\langle\bar{s}g_s\sigma Gs\rangle^2}{18432\pi^4}\int_{x_i}^{1}dx(1-x)^2-\frac{\langle\bar{s}g_s\sigma Gs\rangle^2}{73728\pi^4}\int_{x_i}^{1}dxs(1-x)^2\delta(s-\tilde{m}_c^2)\nonumber\\
&&+\frac{m_sm_c\langle\bar{q}g_s\sigma Gq\rangle\langle\bar{s}g_s\sigma Gs\rangle}{36864\pi^4}\int_{x_i}^{1}dx\frac{1+23x}{x}\delta(s-\tilde{m}_c^2)\, ,
\end{eqnarray}

\begin{eqnarray}
\rho^1_{D^*\Xi}(11)&=&-\frac{m_c\langle\bar{s}s\rangle^2\langle\bar{q}g_s\sigma Gq\rangle}{96\pi^2}\int_{x_i}^{1}dx\delta(s-\tilde{m}_c^2)\nonumber\\
&&-\frac{m_c\langle\bar{s}s\rangle\langle\bar{q}q\rangle\langle\bar{s}g_s\sigma Gs\rangle}{96\pi^2}\int_{x_i}^{1}dx\delta(s-\tilde{m}_c^2)\, ,
\end{eqnarray}

\begin{eqnarray}
\rho^1_{D^*\Xi}(13)&=&-\frac{m_c^3\langle\bar{q}q\rangle\langle\bar{s}s\rangle^2}{432T^4}\langle\frac{\alpha_{s}GG}{\pi}\rangle\int_{x_i}^{1}dx\frac{1-x}{x^3}\delta(s-\tilde{m}_c^2)\nonumber\\
&&+\frac{m_c\langle\bar{q}q\rangle\langle\bar{s}s\rangle^2}{144T^2}\langle\frac{\alpha_{s}GG}{\pi}\rangle\int_{x_i}^{1}dx\frac{1-x}{x^2}\delta(s-\tilde{m}_c^2)\nonumber\\
&&-\frac{m_c\langle\bar{s}s\rangle\langle\bar{q}g_s\sigma Gq\rangle\langle\bar{s}g_s\sigma Gs\rangle}{27648\pi^2T^2}\int_{x_i}^{1}dx\frac{1}{x}\delta(s-\tilde{m}_c^2)\nonumber\\
&&+\frac{m_c\langle\bar{q}q\rangle\langle\bar{s}s\rangle^2}{288T^2}\langle\frac{\alpha_{s}GG}{\pi}\rangle\delta(s-m_c^2)\nonumber\\
&&+\frac{m_c\langle\bar{s}s\rangle\langle\bar{q}g_s\sigma Gq\rangle\langle\bar{s}g_s\sigma Gs\rangle}{384\pi^2T^2}\delta(s-m_c^2)\, ,
\end{eqnarray}

\begin{eqnarray}
\rho^0_{D^*\Xi}(3)&=&\frac{\langle\bar{q}q\rangle}{24576\pi^6}\int_{x_i}^{1}dxx(1-x)^4(5s-\tilde{m}_c^2)(s-\tilde{m}_c^2)^3\, ,
\end{eqnarray}

\begin{eqnarray}
\rho^0_{D^*\Xi}(6)&=&-\frac{m_c\langle\bar{q}q\rangle^2}{128\pi^4}\int_{x_i}^{1}dx(1-x)^2(s-\tilde{m}_c^2)^2\nonumber\\
&&+\frac{m_s\langle\bar{q}q\rangle\langle\bar{s}s\rangle}{128\pi^4}\int_{x_i}^{1}dxx(1-x)^2(-9s+3\tilde{m}_c^2)(s-\tilde{m}_c^2)\, ,
\end{eqnarray}

\begin{eqnarray}
\rho^0_{D^*\Xi}(7)&=&-\frac{m_c^2\langle\bar{q}q\rangle}{18432\pi^4}\langle\frac{\alpha_{s}GG}{\pi}\rangle\int_{x_i}^{1}dx\frac{(1-x)^4}{x^2}(2s-\tilde{m}_c^2)\nonumber\\
&&+\frac{\langle\bar{q}q\rangle}{9216\pi^4}\langle\frac{\alpha_{s}GG}{\pi}\rangle\int_{x_i}^{1}dx(1-x)^2[(x-1)(3\tilde{m}_c^2-5s)+3x(\tilde{m}_c^2-3s)](s-\tilde{m}_c^2)\, ,\nonumber \\
\end{eqnarray}

\begin{eqnarray}
\rho^0_{D^*\Xi}(8)&=&\frac{5m_s\langle\bar{q}q\rangle\langle\bar{s}g_s\sigma Gs\rangle}{192\pi^4}\int_{x_i}^{1}dxx(1-x)(2s-\tilde{m}_c^2)\nonumber\\
&&+\frac{7m_s\langle\bar{s}s\rangle\langle\bar{q}g_s\sigma Gq\rangle}{256\pi^4}\int_{x_i}^{1}dxx(1-x)(2s-\tilde{m}_c^2)\nonumber\\
&&+\frac{m_c\langle\bar{q}q\rangle\langle\bar{q}g_s\sigma Gq\rangle}{128\pi^4}\int_{x_i}^{1}dx(1-x)(s-\tilde{m}_c^2)\, ,
\end{eqnarray}

\begin{eqnarray}
\rho^0_{D^*\Xi}(9)&=&\frac{\langle\bar{q}q\rangle\langle\bar{s}s\rangle^2}{12\pi^2}\int_{x_i}^{1}dxx(1-x)(2s-\tilde{m}_c^2)+\frac{m_sm_c\langle\bar{q}q\rangle^2\langle\bar{s}s\rangle}{8\pi^2}\int_{x_i}^{1}dx\, ,
\end{eqnarray}

\begin{eqnarray}
\rho^0_{D^*\Xi}(10)&=&\frac{m_c\langle\bar{q}g_s\sigma Gq\rangle^2}{36864\pi^4}\int_{x_i}^{1}dx(1-x)+\frac{m_s\langle\bar{q}g_s\sigma Gq\rangle\langle\bar{s}g_s\sigma Gs\rangle}{6144\pi^4}\int_{x_i}^{1}dx(1-45x)\nonumber\\
&&+\frac{m_s\langle\bar{q}g_s\sigma Gq\rangle\langle\bar{s}g_s\sigma Gs\rangle}{18432\pi^4}\int_{x_i}^{1}dxs(1-133x)\delta(s-\tilde{m}_c^2)\nonumber\\
&&+\frac{m_c^3\langle\bar{q}q\rangle^2}{1152\pi^2}\langle\frac{\alpha_{s}GG}{\pi}\rangle\int_{x_i}^{1}dx\frac{(1-x)^2}{x^3}\delta(s-\tilde{m}_c^2)\nonumber\\
&&-\frac{m_c\langle\bar{q}q\rangle^2}{1152\pi^2}\langle\frac{\alpha_{s}GG}{\pi}\rangle\int_{x_i}^{1}dx\frac{3-6x+2x^2}{x^2}\nonumber\\
&&+\frac{m_sm_c^2\langle\bar{s}s\rangle\langle\bar{q}q\rangle}{384\pi^2T^2}\langle\frac{\alpha_{s}GG}{\pi}\rangle\int_{x_i}^{1}dx\frac{s(1-x)^2}{x^2}\delta(s-\tilde{m}_c^2)\nonumber\\
&&-\frac{m_s\langle\bar{s}s\rangle\langle\bar{q}q\rangle}{1152\pi^2}\langle\frac{\alpha_{s}GG}{\pi}\rangle\int_{x_i}^{1}dx(18-11x)\nonumber\\
&&-\frac{m_s\langle\bar{s}s\rangle\langle\bar{q}q\rangle}{1152\pi^2}\langle\frac{\alpha_{s}GG}{\pi}\rangle\int_{x_i}^{1}dxs(6+x)\delta(s-\tilde{m}_c^2)\, ,
\end{eqnarray}

\begin{eqnarray}
\rho^0_{D^*\Xi}(11)&=&-\frac{\langle\bar{q}q\rangle\langle\bar{s}s\rangle\langle\bar{s}g_s\sigma Gs\rangle}{24\pi^2}\int_{x_i}^{1}dxx[1+s\delta(s-\tilde{m}_c^2)]\nonumber\\
&&-\frac{\langle\bar{s}s\rangle^2\langle\bar{q}g_s\sigma Gq\rangle}{48\pi^2}\int_{x_i}^{1}dxx[1+s\delta(s-\tilde{m}_c^2)]\nonumber\\
&&-\frac{5m_sm_c\langle\bar{q}q\rangle^2\langle\bar{s}g_s\sigma Gs\rangle}{144\pi^2}\delta(s-m_c^2)-\frac{13m_sm_c\langle\bar{q}q\rangle\langle\bar{s}s\rangle\langle\bar{q}g_s\sigma Gq\rangle}{192\pi^2}\delta(s-m_c^2)\, ,
\end{eqnarray}

\begin{eqnarray}
\rho^0_{D^*\Xi}(12)&=&-\frac{m_c\langle\bar{q}q\rangle^2\langle\bar{s}s\rangle^2}{9}\delta(s-m_c^2)\, ,
\end{eqnarray}

\begin{eqnarray}
\rho^0_{D^*\Xi}(13)&=&\frac{m_c^2\langle\bar{s}s\rangle^2\langle\bar{q}q\rangle}{216T^2}\langle\frac{\alpha_{s}GG}{\pi}\rangle\int_{x_i}^{1}dx\frac{(1-x)}{x^2}\left(1-\frac{s}{T^2}\right)\delta(s-\tilde{m}_c^2)\nonumber\\
&&+\frac{\langle\bar{s}s\rangle^2\langle\bar{q}q\rangle}{108}\langle\frac{\alpha_{s}GG}{\pi}\rangle\int_{x_i}^{1}dx\left(1+\frac{s}{2T^2}\right)\delta(s-\tilde{m}_c^2)\nonumber\\
&&-\frac{m_sm_c^3\langle\bar{s}s\rangle\langle\bar{q}q\rangle^2}{144T^4}\langle\frac{\alpha_{s}GG}{\pi}\rangle\int_{x_i}^{1}dx\frac{1}{x^3}\delta(s-\tilde{m}_c^2)\nonumber\\
&&+\frac{m_sm_c\langle\bar{s}s\rangle\langle\bar{q}q\rangle^2}{48T^2}\langle\frac{\alpha_{s}GG}{\pi}\rangle\int_{x_i}^{1}dx\frac{1}{x^2}\delta(s-\tilde{m}_c^2)\nonumber\\
&&-\frac{m_sm_c\langle\bar{s}s\rangle\langle\bar{q}g_s\sigma Gq\rangle^2}{27648\pi^2T^2}\int_{x_i}^{1}dx\frac{1}{x}\delta(s-\tilde{m}_c^2)\nonumber\\
&&-\frac{\langle\bar{s}s\rangle\langle\bar{q}g_s\sigma Gq\rangle\langle\bar{s}g_s\sigma Gs\rangle}{6912\pi^2}\int_{x_i}^{1}dx\left(1+\frac{s}{2T^2}\right)\delta(s-\tilde{m}_c^2)\nonumber\\
&&+\frac{5sm_sm_c\langle\bar{s}s\rangle\langle\bar{q}q\rangle^2}{432T^4}\langle\frac{\alpha_{s}GG}{\pi}\rangle\delta(s-m_c^2)+\frac{s\langle\bar{s}s\rangle^2\langle\bar{q}q\rangle}{144T^2}\langle\frac{\alpha_{s}GG}{\pi}\rangle\delta(s-m_c^2)\nonumber\\
&&+\frac{m_sm_c\langle\bar{s}s\rangle\langle\bar{q}g_s\sigma Gq\rangle^2}{55296\pi^2T^2}\left(1+\frac{504s}{T^2}\right)\delta(s-m_c^2)\nonumber\\
&&+\frac{s\langle\bar{q}q\rangle\langle\bar{s}g_s\sigma Gs\rangle^2}{192\pi^2T^2}\delta(s-m_c^2)+\frac{s\langle\bar{s}s\rangle\langle\bar{s}g_s\sigma Gs\rangle\langle\bar{q}g_s\sigma Gq\rangle}{96\pi^2T^2}\delta(s-m_c^2)\nonumber\\
&&+\frac{7sm_sm_c\langle\bar{q}q\rangle\langle\bar{s}g_s\sigma Gs\rangle\langle\bar{q}g_s\sigma Gq\rangle}{384\pi^2T^4} \delta(s-m_c^2)\, .
\end{eqnarray}

{\bf  For the $D^*N$ pentaquark molecular states},
\begin{eqnarray}
\rho_{D^*N}^1(s)&=&\rho_{D^*\Xi}^1(s)\mid_{m_s \to 0, \langle\bar{s}s\rangle \to\langle\bar{q}q\rangle, \langle\bar{s}g_s\sigma Gs\rangle \to\langle\bar{q}g_s\sigma Gq\rangle} \, , \nonumber\\
\rho_{D^*N}^0(s)&=&\rho_{D^*\Xi}^0(s)\mid_{m_s \to 0, \langle\bar{s}s\rangle \to\langle\bar{q}q\rangle, \langle\bar{s}g_s\sigma Gs\rangle \to\langle\bar{q}g_s\sigma Gq\rangle} \, .
\end{eqnarray}

\section*{Acknowledgements}
This  work is supported by National Natural Science Foundation, Grant Number 12175068, and Postgraduate Students Innovative Capacity Foundation of Hebei Education Department, Grant
Number CXZZBS2023146.

\end{document}